\documentclass{aa}  

\usepackage{graphicx}
\usepackage{newtxtext,newtxmath}
\usepackage[T1]{fontenc}

\DeclareRobustCommand{\VAN}[3]{#2}
\let\VANthebibliography\thebibliography
\def\thebibliography{\DeclareRobustCommand{\VAN}[3]{##3}\VANthebibliography}

\usepackage{graphicx}
\usepackage{amsmath}
\usepackage{txfonts}
\usepackage{hyperref}
\usepackage{tablefootnote}
\usepackage{siunitx}
\usepackage{threeparttable}
\hypersetup{colorlinks,linkcolor={blue},citecolor={blue},urlcolor={blue}} 

\usepackage[tableposition=top]{caption}
\usepackage[protrusion=true,expansion]{microtype}
\usepackage{color}
\usepackage{titlesec}

\titlespacing*{\paragraph}{0pt}{0pt}{0.5em}
\usepackage{stfloats}

\begin{document} 

   \title{Relatively young thick disks in star-forming late-type galaxies}

   \author{
    Natascha Sattler\inst{1,2}\thanks{\email{n.sattler@stud.uni-heidelberg.de}}
    \and
    Francesca Pinna\inst{3,4,1}\thanks{\email{francesca.pinna@iac.es}}
    \and
    Sebastien Comer\'on\inst{4,3}
    \and
    Marie Martig\inst{5}
    \and
    Jesus Falc\'on-Barroso\inst{3,4}
    \and
    Ignacio Mart\'in-Navarro\inst{3,4}
    \and
    Nadine Neumayer\inst{1}
    }
   \institute{
   Max Planck Institute for Astronomy, Koenigstuhl 17, D-69117 Heidelberg, Germany
   \and
   Astronomisches Rechen-Institut, Zentrum f\"{u}r Astronomie der Universit\"{a}t Heidelberg, M\"{o}nchhofstra\ss e 12-14, D-69120 Heidelberg, Germany 
   \and
   Instituto de Astrofísica de Canarias, Calle Vía Láctea s/n, E-38205 La Laguna, Tenerife, Spain
   \and 
   Departamento de Astrofísica, Universidad de La Laguna, Av. del Astrofísico Francisco Sánchez s/n, E-38206, La Laguna, Tenerife, Spain
   \and
   Astrophysics Research Institute, Liverpool John Moores University, 146 Brownlow Hill, Liverpool, L3 5RF, UK
   }

  \abstract
  {}{We trace the evolution of eight edge-on star-forming disk galaxies by analyzing stellar population properties of their thin and thick disks. These galaxies have relatively low stellar masses (4~$\times$~10$^9$ to 6~$\times$~10$^{10}$~$M_{\odot}$). }
  {We used Multi-Unit Spectroscopic Explorer (MUSE) observations and a full-spectrum fitting to produce spatially resolved maps of the ages, metallicities, and [Mg/Fe] abundances, and we extracted the star formation histories of the stellar disks.}
  {Our maps show thick disks that are older, more metal-poor, and more [Mg/Fe]-enhanced than thin disks on average.
  The age differences between thin and thick disks are small (about 2~Gyr), however, and the thick disks are younger than previously observed in more massive and more quiescent galaxies. The thin and thick disks both show mostly sub-solar metallicities, and the vertical metallicity gradient is milder than previously observed in similar studies. The [Mg/Fe] differences between thick and thin disks are not sharp. 
  The star formation histories of thick disks extend to recent times, although most of the mass in young stars was formed in thin disks.} 
  {Our findings show thick disks that are different from the old quiescent thick disks that were previously observed in galaxies with different morphologies and/or masses. 
  We propose that the thick disks in these galaxies did not form quickly at high redshift, but formed slowly over an extended time. 
  The thin disks also formed slowly, but a larger mass fraction was created at very recent times.
  } 


   \maketitle

\section{Introduction}
\label{sec:introduction}

A galaxy disk can have multiple disk components with different thicknesses. 
Two components usually form that are called thin and thick disks \citep{burstein_1979, gilmore_1983, yoachim_2006, comeron_2018}. 
The former have a higher surface brightness and dominate the midplane region, while the latter are not as bright and dominate at larger heights. 
The vertical structure of disks can be best studied when they are seen edge-on.
Thin and thick disks in external galaxies are usually defined geometrically by selecting regions at different heights from the midplane \citep{yoachim_2006}.
Thick disks usually have higher velocity dispersions and mostly show older, more metal-poor and $\alpha$-enhanced stellar populations than thin disks \citep[e.g.,][]{yoachim_2008b, comeron_2015, kasparova_2016, pinna_2019b, pinna_2019a, scott_2021, martig_2021, sattler_2023, pinna_2024}. 
Furthermore, in galaxies with a circular velocity > 90~km~s$^{-1}$, they are typically less massive than thin disks, but in galaxies with a circular velocity < 90~km~s$^{-1}$, the mass ratios of thin and thick disks can be lower than one \citep{yoachim_2006, comeron_2011, comeron_2012, comeron_2019}.
Based on this, thin and thick disks can also be identified by their kinematics \citep[e.g.,][]{Morrison1990, Bensby2003, Haywood2013, Collins2011, Vieira2023}, stellar ages \citep[e.g.,][]{Haywood2013, Xiang2015}, and chemical compositions \citep[e.g.,][]{fuhrmann1998, bovy2012}.
\\\\
The differences in the stellar kinematics and populations indicate the different histories of thick and thin disks.
The formation of thick disks can be explained by multiple different processes.
They can be born already thick through gas-rich mergers \citep{brook_2004}, during which the stars in the thick disk are formed in a turbulent phase at high redshift.
They can form through instabilities in the disk \citep{bournaud_2009}, which leads to strong stellar scattering. 
Instabilities arising from internal processes would result in a thick disk with a constant scale height at all galactocentric radii \citep{bournaud_2009}.
In addition, thick disks may form from a previously settled thin disk that is dynamically heated by minor mergers \citep{quinn_1993}. 
This scenario leads to a flaring thick disk \citep{bournaud_2009}.
Furthermore, the accretion of stars from satellite galaxies can build a thick disk structure \citep{abadi_2003}.
The stellar accretion might leave retrograde-moving stars in the disk, which can contribute to a high velocity dispersion \citep{yoachim_2008a}.\\
Thin disks, however, are thought to form mainly in-situ.
Thereby, large amounts of gas can be provided by gas-rich galaxy mergers or gas accretion along the filaments of the intergalactic medium, which can fuel an extended star formation with a long chemical enrichment to metal-rich and $\alpha$-poor stars \citep{gallart_2019, martig_2021, conroy_2022}.
An extended chemical enrichment only occurs after subsequent generations of stars pollute the interstellar medium with their metals. The first high-mass stars, which originally formed from metal-poor gas \citep{mo_2010}, exploded as core-collapse supernovae and left some $\alpha$-elements (e.g., magnesium and oxygen) in the interstellar medium \citep[e.g.,][]{worthey_1992, mo_2010, peletier_2013}. 
From this more $\alpha$-enriched gas, new generations of stars are formed.
The enrichment with heavy elements (e.g., iron and nickel) mostly occurs when stars explode as type~Ia supernovae, whose progenitors have a much longer lifetime than those of core-collapse supernovae.
This causes a delay in the production of iron with respect to $\alpha$-elements.
Therefore, unless they form from recently accreted pristine gas, the youngest stars, which are primarily located in the thin disk, have a lower [$\alpha$/Fe] abundance because they form from slowly evolved gas that contains a larger amount of heavier metals. This makes them generally more metal-rich.
\\
All these different formation mechanisms of thin and thick disks can be combined to explain the build-up of observed galaxies \citep[e.g.,][]{comeron_2016, pinna_2019b, pinna_2019a, santucci_2020, martig_2021}.
Several studies proposed that galaxy formation occurs in two phases: In a first rapid early phase during which stars are formed in-situ in the galaxies themselves, and in a second extended phase in which ex-situ stars are accreted onto the galaxies or in which in-situ star formation, fueled by mergers, occurs \citep[e.g.,][]{brook_2004, Oser2010, Obreja2012, Dominguez-Tenreiro2017, yu_2021, pinna_2024}. 
This can create a [$\alpha$/Fe]-dichotomy \citep[e.g.,][]{Grand2018}.
Thereby, stellar migration might also play an important role in the formation of a thick disk because during this first violent phase of galaxy formation, stellar orbits tend to diffuse and migrate \citep{Halle2018}, while in later phases, stars also migrate from the inner disk toward larger radii \citep{Hayden2018}.
\citet{Comeron2014} suggested that the thin disk forms after the build-up of the dynamically hot component (which includes the thick disk and a central mass concentration).
The dynamically cold component (which includes the thin disk and molecular gas) has enough gas especially in low-mass galaxies to produce a thin disk in the future that is as massive, in relative terms, as current thin disks in high-mass galaxies that already exhausted most of their gas.
Therefore, even when the mass ratio of thick and thin disks varies as a function of galaxy mass, the mass ratio of the dynamically hot and cold components stays roughly constant.
Moreover, cosmological simulations of Milky Way-sized galaxies \citep{martig_2012, grand_2017} were also used to study the growth of disks.
They show that a thin disk is often present since the earliest evolutionary phase and grows continuously inside-out \citep{pinna_2024}. 
This inside-out process is thought to describe the mass assembly of galaxies with a mass higher than 10$^{10.5}$~$M_{\odot}$ \citep[e.g.,][]{perez_2013, pan_2015} and can be traced by a negative radial age gradient, with older stars in the central region and younger stars in the outskirts \citep{perez_2013}.
In contrast, galaxies with a mass lower than 10$^{10}$~$M_{\odot}$ are proposed to form from an outside-in process that would leave younger stars in the center and older stars in the outskirts. This can thus be traced by a positive radial age gradient \citep{zhang_2012, pan_2015}.
\\\\
Distinct thin and thick disks have been identified in numerous galaxies.
For example, \citet{yoachim_2006} reported that 32 out of 34 edge-on galaxies host a thick disk, while \citet{comeron_2018} separated 124 out of 141 galaxies into two large-scale disk structures.
The Milky Way is also known to have multiple disk structures \citep{gilmore_1983}.
Being in the Milky Way allows us to gain insights into the thin and thick disk formation from the inside, for example, through the APOGEE (Apache Point Observatory Galactic Evolution Experiment) and Gaia projects \citep{Majewski2015, gaia-collaboration_2022, Vieira2023}.
Using projected tangential velocities, \citet{Vieira2023} obtained thin and thick disk scale heights of 279.76~$\pm$~12.49~pc and 797.23~$\pm$~12.34~pc, respectively.
In the solar neighborhood, older more metal-poor and $\alpha$-enhanced stars are found at larger distances from the midplane \citep{gilmore_1985, ivezic_2008, schlesinger_2012, casagrande_2016}.
Radial gradients in age and [$\alpha$/Fe] are observed in the geometric thick disk of the Milky Way \citep{minchev_2015, martig_2016, hayden_2017, queiroz_2020, gaia-collaboration_2022}. 
The outer thick disk is dominated by younger stars (about 5~Gyr) that are $\alpha$-poor, whereas the inner thick disk is older (about 9~Gyr) and more enhanced in $\alpha$-elements.
These radial gradients may result from different formation processes for the inner and outer thick disk, where the inner thick disk is formed already thick at high redshift, while the outer thick disk grows continuously and external perturbations bring young and $\alpha$-poor stars into the outer thick disk \citep{minchev_2015}.
Otherwise, radial gradients may also emerge from the flaring of the outer region of the thin disk, as was shown for lenticulars \citep[e.g.,][]{pinna_2019b, pinna_2019a} and Milky Way-mass spiral galaxies \citep{pinna_2024}. \\

Due to the limited amount of spectroscopic data, it is still unclear whether the processes described above truly describe the formation of the Milky Way and if they can also be applied to external galaxies. 
A few long-slit studies analyzed the stellar kinematics and stellar populations for the thin and thick disks of edge-on galaxies \citep[e.g.,][]{yoachim_2008a, yoachim_2008b, katkov_2019, kasparova_2016, kasparova_2020}.
\citet{yoachim_2008b,yoachim_2008a} provided stellar and gaseous kinematic measurements of nine edge-on late-type galaxies, as well as luminosity-weighted stellar ages and metallicities for their thin and thick disks.
The kinematics of the thick disks in their sample favored an accretion origin for thick disk stars.
Moreover, they possess an old stellar population between 4~and~10~Gyr, while the thin disks are overall younger with a strong negative radial age gradient.
\citet{kasparova_2016} obtained deep spectra of thick disks in three edge-on S0 galaxies.
Two of these galaxies possess stellar populations with similar properties ([Fe/H]~$\sim$~-0.2 to 0.0~dex and ages of about 4-5~Gyr) for both disk components, while the third galaxy shows only one old (thick) disk component.
\\\\
However, with the use of integral-field spectroscopy, stellar properties such as kinematics and populations can be mapped spatially in a continuous way that includes the transition region between thin and thick disks, and provide detailed insights into the structure of extragalactic disks.
Several studies have already been conducted on bright edge-on S0 galaxies \citep{comeron_2016, guerou_2016, pinna_2019b, pinna_2019a}. 
The quiescent galaxies studied by \citet{comeron_2016, pinna_2019b, pinna_2019a} showed very old and metal-poor thick disks (which are also $\alpha$-enhanced according to \citealt{pinna_2019b, pinna_2019a}), suggesting their short and fast formation at high redshift.
Moreover, \citet{pinna_2019b, pinna_2019a} detected a subdominant but significant and potentially accreted component in the stellar populations of three lenticular galaxies in the Fornax cluster. 
The study of faint thick disks in spiral galaxies is more challenging because their surface brightness is low and their large amounts of gas (producing striking emission lines) and dust hinder a detailed analysis. 
The first attempt to study spiral galaxies was made with VIMOS (VIsible MultiObject Spectrograph) by \citet{comeron_2015}, but the spatial resolution was poor.
The authors proposed that the old thick disk of ESO\,533-4 was born already dynamically hot in a turbulent disk or formed from dynamical heating.
\\\\

\begin{figure*}
    \centering
    \includegraphics[width=0.8\textwidth]{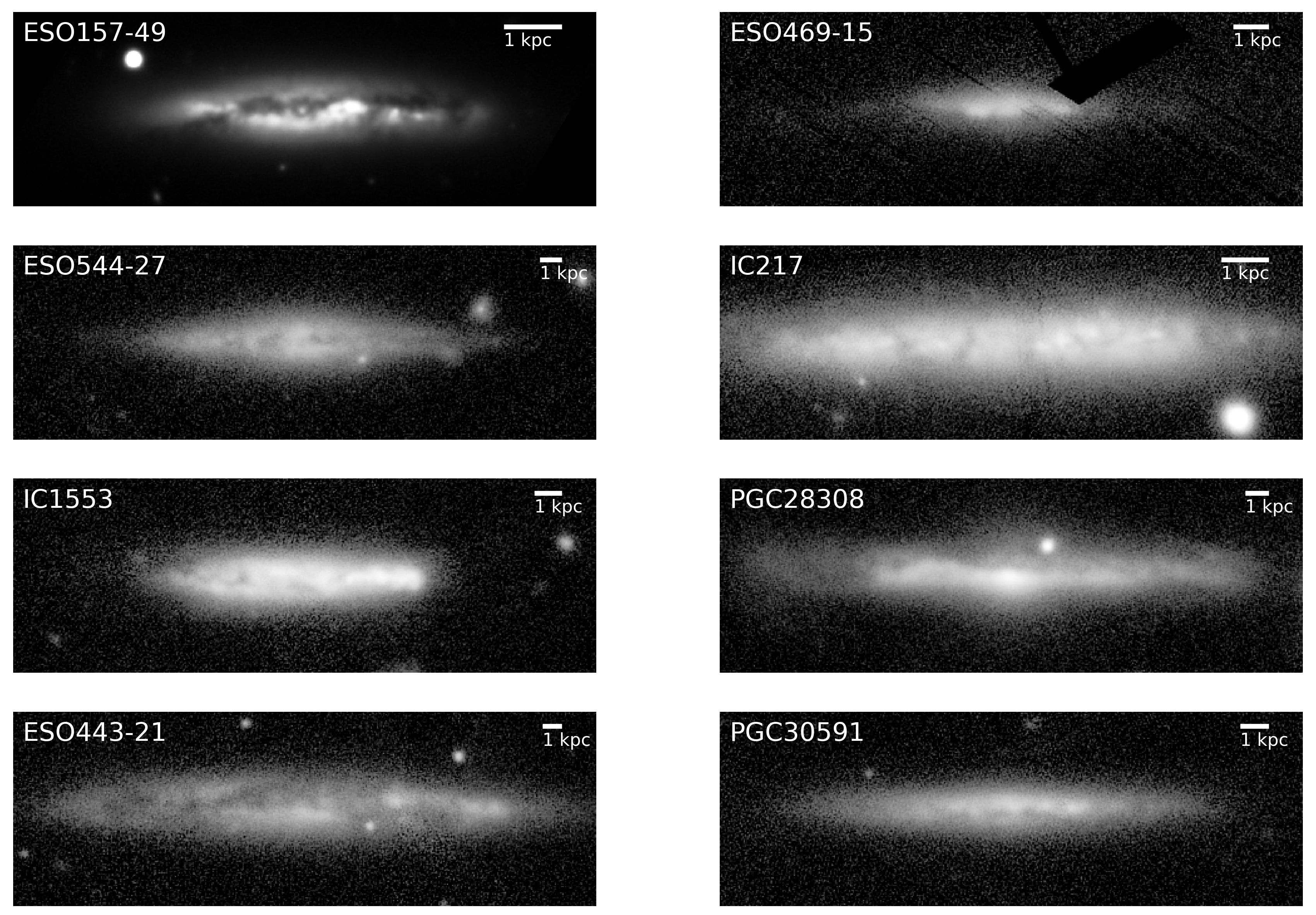} 
    \caption{\textit{g}-band images of the full sample. The image of ESO\,157-49 is taken from the Dark Energy Survey DR2
    \citep{DES_2018, DES_2021}, while the other images are from Pan-STARRS1
    \citep{panstarrs_2016, panstarrs_2020a, panstarrs_2020b, panstarrs_2020c, panstarrs_2020d, panstarrs_2020e}.}
    \label{fig:g_band}
\end{figure*}

Spatially resolved maps of the stellar and gas kinematics with a better resolution were later achieved by \citet{comeron_2019} using MUSE (Multi-Unit Spectroscopic Explorer) for eight late-type edge-on star-forming field disk galaxies.
This study included five galaxies with a distinct thick disk in the observed region and disfavored the accretion of stars from satellite galaxies as the main formation mechanism of thick disks. 
\citet{rautio_2022} examined the gas content for these five galaxies in more detail. The vertical gas velocity gradients or asymmetries in the gas kinematics indicated gas accretion (see Sect.~\ref{sec:sample} for more details).
\citet{martig_2021}, \citet{scott_2021}, and \citet{sattler_2023} studied the stellar populations of the spiral galaxies NGC\,5746, UGC\,10738, and NGC\,3501. 
The study of the massive disk galaxy NGC\,5746 \citep{martig_2021} supported the idea (previously proposed by \citealt{pinna_2019b, pinna_2019a}) that the thick disk might be born already thick in an early phase of evolution, and that an additional large fraction of stars (about 30~\%) was accreted to the thick disk during a merger event about 8~Gyr ago.
For NGC\,3501, a star-forming galaxy with apparently no morphological distinct thick and thin disks in the observed region, \citet{sattler_2023} found indications of the recent birth and current inside-out growth of a thin disk that is embedded in a preexisting thicker disk.

\begin{table*}
    \centering
    \caption{Properties of the sample galaxies.}
    \begin{tabular}{ccccccccccc}
        \hline
        \\
        Galaxy & Hubble  & $d$ $^{\left(2\right)}$& log($M_*$)$^{\left(3\right)}$ & $r_{25}$ $^{\left(4\right)}$& $v_{\text{c}}$ $^{\left(5\right)}$& Disks $^{\left(6\right)}$& $z_{\text{c1}}$ $^{\left(7\right)}$& $M_{\text{t}}$ $^{\left(8\right)}$& $M_{\text{T}}$ $^{\left(9\right)}$& SFR $^{\left(10\right)}$\\
        name & type $^{{\left(1\right)}}$ & [Mpc] & [$M_{\odot}$] & [$\si{\arcsecond}$] & [km s$^{-1}$] & & [$\si{\arcsecond}$] / [kpc] & \multicolumn{2}{c}{[$10^9 M_{\odot}$]} & [$M_{\odot}$/yr]\\
        \hline
        \\
        \hfill{} ESO\,157-49 & Sc & 17.3 & \ 9.81 & 52.1 & 107 & t + T & 6.1 / 0.5 & 2.54 & 1.21 & 0.19\\
        \hfill{} ESO\,469-15 & Sb & 28.3 & \ 9.58 & 55.9 & \ 83 & t + T & 4.8 / 0.7 & 2.82 & 1.94 & 0.20\\
        \hfill{} ESO\,544-27 & Sb & 45.9 & \ 9.84 & 46.5 & 129 & t + T & 4.0 / 0.9 & 7.56 & 3.57 & 0.13\\
        \hfill{} IC\,217 & Scd & 21.1 & \ 9.75 & 59.9 & 115 & t + T & 5.4 / 0.6 & 2.46 & 1.72 & 0.25\\
        \hfill{} IC\,1553 & Irr & 36.5 & \ 9.90 & 40.5 & 142 & t + T & 5.0 / 0.9 & 8.63 & 4.00 & 0.96\\
        \hfill{} PGC\,28308 & Scd & 43.1 & 10.81 & 59.9 & 130 & t*+ T & - & - & - & (0.51)\\
        \hfill{} ESO\,443-21 & Scd & 52.2 & 10.34 & 36.9 & 196 & - & - & - & - & (2.40)\\
        \hfill{} PGC\,30591 & Sd & 35.5 & 10.21 & 45.4 & \ 97 & - & - & - & - & (0.29)\\
        \hline
        \\
    \end{tabular}
    \label{tab:prop}
        \begin{tablenotes}
            \item \textbf{Notes:} (1) Morphological types from \citet{gonzalez_2024}.
            (2) Distances ($d$) from \citet{tully_2008, tully_2016}. (3) Stellar masses ($M_*$) from \citet{munoz_2015}. (4) Radii ($r_{25}$) from HyperLeda \citep{makarov_2014}. (5) Circular velocities ($v_{\text{c}}$) from \citet{comeron_2019}. (6) Disk structures (t: thin disk, T: thick disk, *: thin disk dominates at all heights) from \citet{comeron_2018}. (7) Distance from the midplane beyond which the thick disk is dominant ($z_{\text{c1}}$). (8) and (9): Masses of the thin ($M_{\text{t}}$) and thick disks ($M_{\text{T}}$). (10) Star formation rates from \citet{rautio_2022, Rautio2024}. The values in brackets have high uncertainties.
        \end{tablenotes}
\end{table*}

With this paper, we wish to extend the knowledge about the thin and thick disk formation in spiral galaxies by analyzing the stellar populations of the eight late-type galaxies for which the kinematics were published by \citet{comeron_2019} in great detail.
The paper is structured as follows:
First, we give an overview of the sample in Sect.~\ref{sec:sample}, which is followed by a description of the observations and data reduction in Sect.~\ref{sec:obs}.
Next, the different analytical steps are explained in Sect.~\ref{sec:methods}. 
The results are presented in Sect.~\ref{sec:results} and are discussed in Sect.~\ref{sec:discussion}.
The conclusions are then given in Sect.~\ref{sec:conclusion}.

\section{The sample}
\label{sec:sample}
The sample consists of the eight edge-on disk galaxies presented in \citet{comeron_2019}, who later released their reduced data cubes (see Sect.~\ref{sec:obs}). 
We show \textit{g}-band images of the target galaxies in Fig.~\ref{fig:g_band}, whereas relevant properties of the sample are indicated in Tab.~\ref{tab:prop}. 
Galaxies in the sample have morphological types from Sb to Irr \citep{gonzalez_2024} and masses between 4~$\times$~10$^9$ and 6~$\times$~10$^{10}$~$M_{\odot}$ \citep{munoz_2015}.

\subsection{Surface brightness profiles}
\label{sub:SB_profiles}
The vertical and radial surface brightness profiles of these galaxies were previously analyzed by \citet{comeron_2018} using Spitzer S$^4$G \citep{sheth_2010} images.
For each galaxy, they performed fits on the vertical surface brightness profiles using the solution of the hydrostatic equilibrium equation under the assumption of two vertically isothermal components (thin and thick disk).
Furthermore, central mass concentrations (CMCs) were identified in ESO\,157-49, IC\,1553, PGC\,28308, and ESO\,443-21.
These can be classical bulges, or boxy/peanut/X-shaped features. In all cases, they were assumed to have a spherical shape and were fit with a S\'ersic function.
To fit surface brightness profiles of galaxies hosting this CMC, \citet{comeron_2018} used three components (CMC, thin and thick disks). 
Based on those fits, they separated their sample of 141 edge-on galaxies into two subsets:
galaxies having a thin and thick disk (here, the vertical profiles of the thin and thick disk dominate at different heights) and galaxies that do not show two morphologically distinct disks (here, the vertical profile fit of only one component dominates at all heights). 
Using this definition of the two subsets, the galaxies ESO\,157-49, ESO\,469-15, ESO\,544-27, IC\,217, and IC\,1553 show two distinct thick and thin disk components that dominate over different heights. 
For this subsample, we used in the rest of the paper a geometric definition of the thick and thin disks, based on this morphological decomposition where the thick disk starts to dominate over the thin disk at a distance from the midplane equal to $z_{c1}$ (Tab.~\ref{tab:prop}). 
For ESO\,443-21 and PGC\,30591, the thick disk component dominates the light along all heights, which is why they were classified by \citet{comeron_2018} as galaxies not showing two distinct disk components.
The last galaxy PGC\,28308 was found to have distinct and well-defined thin and thick disks. 
However, the thin disk dominates the vertical surface brightness profile at all heights. 
Because of this, we assign here PGC\,28308 to the subsample of galaxies that do not have two distinct stellar disks.

\subsection{Kinematics}
\citet{comeron_2019} extracted and analyzed the kinematics of the stellar and gas content from the same data used here (Sect.~\ref{sec:obs}).
They obtained spatially resolved maps of stellar velocities and velocity dispersions, together with gas velocities for these galaxies.
From the gas velocity maps, they extracted the midplane gas rotation curves and fit these to estimate the circular velocities, which are listed in Tab.~\ref{tab:prop}. 
On average, they detected lower velocity dispersions for the thin disk (with values lower than 20~km~s$^{-1}$) than for the thick disk (40-60~km~s$^{-1}$). 
Also, the galaxies without distinct thick and thin disk components show this pattern of lower velocity dispersion for the midplane compared to regions at larger heights.
This pattern is a consequence of asymmetric drift of the gas and is found in many observed galaxies, as pointed out in Sect.~\ref{sec:introduction}.
For PGC\,28308, the highest velocity dispersion of around 70~km~s$^{-1}$ is present in a spheroidal central structure, which results from a central mass concentration hosting 10~\% of PGC\,28308's baryonic mass \citep{comeron_2019}.

\subsection{Gaseous and stellar properties}
Using the same MUSE data analyzed by \citet{comeron_2019} and in this paper, together with deep narrow-band H$\alpha$ images, \citet{rautio_2022, Rautio2024} analyzed the ionized-gas content for the full sample of eight galaxies. 
They found significant complexity in the extraplanar diffuse ionized gas (eDIG) and H\,\textsc{ii} regions. 
Moreover, they calculated the star formation rates (SFR) for those galaxies and divided their subsample into star-forming and green valley galaxies. 
The more quiescent galaxy ESO\,544-27 has a SFR~=~0.13~$M_{\odot}$/yr, placing it into the green valley, while also PGC\,28308 (SFR~=~0.51~$M_{\odot}$/yr) and PGC\,30591 (SFR~=~0.29~$M_{\odot}$/yr) fall within this regime.
The other galaxies, ESO\,157-49 (SFR~=~0.19~$M_{\odot}$/yr), ESO\,469-15 (SFR~=~0.20~$M_{\odot}$/yr) and IC\,217 (SFR~=~0.25~$M_{\odot}$/yr) lay in the star-forming regime, very close to the line separating it from the green valley.
IC\,1553 (SFR~=~0.96~$M_{\odot}$/yr) and ESO\,443-21 (SFR~=~2.40~$M_{\odot}$/yr) have the highest SFR of the sample and seem to fall well within the star-forming regime.
However, the SFR measurements for ESO~443-21, PGC\,28308, and PGC\,30591 have high uncertainties due to their incomplete spatial coverage in the MUSE field-of-view and should be handled with caution.
Moreover, \citet{rautio_2022} found signs of gas accretion in several galaxies of the sample. 
Furthermore, using eDIG kinematics, Baldwin-Phillips-Terlevich (BPT, \citealt{baldwin_1981}) diagrams and H$\alpha$ intensity maps \citet{rautio_2022, Rautio2024} discussed the ionization sources of the galaxies. 
Another study by \citet{gonzalez_2024} analyzed the gas content in the halos of this same sample of eight galaxies using the same MUSE data by \citet{comeron_2019}.
They suggested the halo gas emission in general originates from feedback-induced shocks from strong star formation in the disk and found hints of an ``intricate dynamical heating structure'' at large distances from the midplane.
In a study by \citet{schwarzkopf_2000}, edge-on disk galaxies were divided into non-interacting and interacting/minor-merging candidates. 
There, we only find one of our eight sample galaxies, ESO\,443-21, which falls into the interacting/minor-merging candidate sample. 
Moreover, in \citet{zaragoza-cardiel_2020}, stellar population properties were measured over several 500~pc sized regions across the whole galaxy disks of ESO\,443-21, IC\,217, IC\,1553, PGC\,28308 and PGC\,30591 and the SFRs for ``recent'' (0.01~Gyr) and ``past'' (0.57~Gyr) age bins were derived.
The details of the above-mentioned and further studies for each galaxy are given in the following section.

\subsection{Summary for individual galaxies}

\paragraph{ESO\,157-49:}The modeling of the kinematics of the ionized gas by \citet{rautio_2022} shows strong evidence for gas accretion, while also the highest H$\alpha$ intensity is observed on the western side of the thin disk approximately 20~arcsec from the center. 
Asymmetries in the eDIG of the halos were also observed by \citet{gonzalez_2024}.
Since ESO\,157-49 has a nearby companion dwarf (ESO\,157-48) with a projected distance of 14.2~kpc, this companion might be a gas accretion source.
Further studies of ESO\,157-49 \citep{keeney_2013} showed signs of a galactic fountain of recycled gas in the kinematics of major axis clouds, while clouds along the minor axes were constrained to be outflowing gas.
Outflows for ESO\,157-49 were also suggested in \citet{lopez_2020}. 
\paragraph{ESO\,469-15:}This galaxy contains discrete extraplanar H\,\textsc{ii} regions at larger heights (3~kpc) above the midplane (in the region dominated by the thick disk) and also more outward (8~kpc) in the radial direction from the center \citep{rautio_2022}. 
These H\,\textsc{ii} regions, with lower rotation velocities and larger velocity dispersions than the ones near the midplane, may cause a more enhanced vertical gradient in the gas rotation velocity than the diffuse gas itself, pointing to a different origin for those H\,\textsc{ii} regions from the eDIG.
\paragraph{ESO\,544-27:}Besides having the lowest star formation rate, this galaxy shows an asymmetric vertical gas-velocity gradient on the southwestern side of the midplane, which can be explained by gas accretion \citep{rautio_2022}.
However, \citet{gonzalez_2024} find a homogeneous distribution of H\,\textsc{ii} regions in the halo and no strong asymmetries along the major axis.
Besides being the most quiescent galaxy, it was also suggested by \citet{rautio_2022} that ESO\,544-27 has older stellar populations than the other galaxies.
While the ionization by in-situ evolved stars is insignificant for most of the galaxies in our sample, it may be able to explain enhanced high-ionization lines in the eDIG of the green valley galaxy ESO\,544-27.
\citet{Somawanshi2024} performed a stellar population analysis, following a similar approach as in \citet{pinna_2019b, pinna_2019a}, \citet{sattler_2023}, and this paper, of the galaxy ESO\,544-27. 
They found signs of recent star formation in the thick disk, and they proposed that the thin and thick disks of ESO\,544-27 formed in-situ.
The comparison of their results with the ones presented here is discussed in Sect.~\ref{subsec:discuss_formation}. 
\paragraph{IC\,217:}This galaxy shows an asymmetric eDIG morphology with very pronounced filaments on the southwestern half and minimal emission on the northeastern one.
A vertical gas velocity gradient could not be confirmed with the MUSE data in \citet{rautio_2022}, since the region with high H$\alpha$ emission is not fully covered by the data.
However, asymmetries in the gas hint at gas accretion.
Also, IC\,217 shows a stronger SFR in the past when compared with more recent times \citep{zaragoza-cardiel_2020}.
\paragraph{IC\,1553:}The most strongly star-forming galaxy, the irregular galaxy IC\,1553 (Tab.~\ref{tab:prop}), shows a rich and asymmetric eDIG morphology with numerous filaments, an extraplanar H\,\textsc{ii} region on the western side, and very pronounced eDIG emission on the southern half \citep{rautio_2022}. 
This galaxy shows a negative vertical gas velocity gradient on the side with high H$\alpha$ emission, suggesting an accretion origin for the gas.
This same high emission region coincides with a conical-shaped structure above and below the midplane in various emission line ratio maps (see also \citealt{gonzalez_2024}).
Also in this galaxy, most regions show stronger past star formation, but there is still a significant number of regions with more star formation in recent ages \citep{zaragoza-cardiel_2020}.
\paragraph{PGC\,28308:}\citet{Tully_2009} found that the H\,\textsc{i} line profile for PGC\,28308 is antisymmetric, which could indicate disturbed outskirts. 
Moreover, concentrated eDIG above the center of this galaxy could hint at ionization by outflow-driven shocks \citet{Rautio2024}.
Also, PGC\,28308 forms a pair with the galaxy MCG-02-25-019 in the HIPASS catalog \citep{Meyer_2004} that has a projected distance of 59~kpc \citep{rautio_2022}.
Further, PGC\,28308 shows a stronger SFR in the past when compared with more recent times \citep{zaragoza-cardiel_2020}
\paragraph{ESO\,443-21:}Most regions in this galaxy have a strong recent star formation \citep{zaragoza-cardiel_2020} which might be due to the interaction with a satellite \citep{schwarzkopf_2000}.
\paragraph{PGC\,30591:}This galaxy shows a stronger SFR in the past when compared with more recent times \citep{zaragoza-cardiel_2020}.

\section{Observations and data reduction}
\label{sec:obs}

We used the archival data published by \citet{comeron_2019}. Observations were taken using MUSE \citep{bacon_2010} at the Unit Telescope~4 of the VLT (Very Large Telescope) in Paranal.
MUSE is a panoramic integral-field spectrograph with a field-of-view of 1$'$~$\times$~1$'$ in the Wide-Field Mode (WFM) and a pixel size of 0.2$''$~$\times$~0.2$''$.
It can observe in a wavelength range of 4750 to 9300~$\si{\angstrom}$ with a spectral resolution R~=~2000 at the blue end and R~=~4000 at the red end \citep{bacon_2017}.\\
The observations correspond to the ESO programs 097.B-0041 and 096.B-0054\footnote{\url{https://archive.eso.org/scienceportal/home?data_collection=096.B-0054}} (P.I. Sebastien Comer\'{o}n) released on 25th September 2019\footnote{\url{https://www.eso.org/sci/publications/announcements/sciann17233.html
%http://research.iac.es/galeria/sebastiencomeron/MUSE.html
}} \citep{comeron_2019} and were carried out between December 2015 and August 2016. 
For seven out of the eight galaxies, four on-target exposures of 2624~s each were executed, but only three on-target exposures of the same time were taken for IC\,217, resulting in a lower total exposure time for this galaxy.
The single exposures were all centered at the same position with a rotation of 90$^\circ$ between each other. 
\\\\
For the data reduction, \citet{comeron_2019} used the MUSE pipeline version 1.6.2 \citep{weilbacher_2012} in the \textsc{reflex} environment \citep{freudling_2013}.
The different exposures of each galaxy were combined to get a single data cube for each galaxy. 
Off-target sky frames with an exposure time of 240~s were also taken, but they were not used in the later data reduction because of large sky variations between on- and off-target exposures coming from the relatively long on-target exposures.
Nevertheless, \citet{comeron_2019} accounted for the sky by modeling it from regions in the on-target exposures that have a larger distance from the galaxy midplane.
Following that, the combined cubes were cleaned from the sky residuals with \textsc{zap} version 2.1 \citep{soto_2016}.

\section{Methods}
\label{sec:methods}

Here, we describe the methods that we applied to the data. 
While this paper focuses on the stellar populations of our galaxy sample, whose analysis is described in Sect.~\ref{sub:sp_analys}, some previous steps were necessary to prepare the data and are described in Sect.~\ref{sub:prep}. 
The stellar population models we used in all these steps are described in Sect.~\ref{subsec:models}.

\subsection{Stellar population models}
\label{subsec:models}
To fit the spectra of the MUSE data cubes as described in the next sections, a spectral library of stellar models needs to be used.
We used the \citet{vazdekis_2015} MILES SSP (Single-Stellar Population) models in this analysis.
These models have a spectral resolution of 2.51~$\si{\angstrom}$ \citep{falcon-barroso_2011} and cover the wavelength range from 3540 to 7410~$\si{\angstrom}$.
After testing different subsets of models (see App.~\ref{appendix:pop}), we decided to use the full range of MILES SSP models, including:
\begin{itemize}
    \item 12 metallicity [M/H] values from -2.27~dex up to 0.40~dex
    \item 53 age values from 0.03 to 14~Gyr
    \item 2 [$\alpha$/Fe] values: 0.0~dex (solar abundance) and 0.4~dex (super-solar abundance)
\end{itemize}
This adds up to 1272 models in total, where each model is associated with a combination of age, metallicity, and [$\alpha$/Fe] abundance.
As we are focusing on the Mgb region, we use [Mg/Fe] as a tracer of [$\alpha$/Fe].

\subsection{Data cube preparation} 
\label{sub:prep}

\subsubsection{Voronoi binning}
\label{subsec:binning}
As the first step, we cut the data cubes of each galaxy to a specific spatial range to exclude bad pixels at the top and bottom edges of the pointings. 
Also, we masked prominent foreground objects that would influence further analysis and results.\\
After that, we binned each data cube with the Voronoi binning (\textsc{vorbin}) method by \citet{cappellari_2003}.
This method combines different pixels of given integral-field spectroscopy data with an adaptive spatial binning to reach a target-S/N (Signal-to-Noise Ratio) per Voronoi bin while preserving the maximum possible spatial resolution of the data.
We performed the \textsc{vorbin} method on the whole sample using a S/N-threshold of 1, meaning that only pixels with a S/N larger than the S/N-threshold are used for the combination into the Voronoi bin. 
We chose this relatively low S/N-threshold to cover the fainter region of the thick disk.
However, since for IC\,217 the exposure time was relatively short (see Sect.~\ref{sec:obs}), the thick disk of this galaxy is not well covered by our Voronoi bins.\\
We first tested a target-S/N~=~40, but stellar populations of ESO\,469-15 and IC\,1553 showed very large uncertainties (as calculated according to Sect.~\ref{subsec:MC_simulations}), especially in age and metallicity.
We decided to increase the target-S/N to 60 for all galaxies, after proving that the uncertainties in the stellar populations scaled down.
Also, the emission line fits (described in Sect.~\ref{subsec:ppxf_gas}) of ESO\,469-15 improved with the higher target-S/N Voronoi binning.  
For IC\,1553, the target-S/N~=~60 Voronoi binning did not improve significantly the quality of the fits and the uncertainties, so we increased the target-S/N to 100 for this galaxy. 
A summary of the used target-S/N for each galaxy can be found in Tab.~\ref{tab:settings}.\\
Lastly, the spectra of each bin were cut to the wavelength range between 4750 to 5500~$\si{\angstrom}$, to avoid regions at longer wavelengths where the sky subtraction might have left residuals, and to also exclude regions that are not relevant for the analysis here as they do not contain $\alpha$-element or age-sensitive features.

\subsubsection{Gas emission line fitting}
\label{subsec:ppxf_gas}
We fit the gas emission to subtract it from the observed spectrum. We needed to achieve an emission-cleaned spectrum to fit the stellar populations without having to mask the regions of important age-sensitive features, like the H$\beta$ $\lambda$4861 absorption. 

For the fitting of the emission lines, we first tested \textsc{gandalf} written by \citet{sarzi_2006} and previously used by \citet{sattler_2023}.
But when looking at the fits, we noticed that \textsc{gandalf} produced large residuals with wiggles for very strong emission lines, for H$\beta$ $\lambda$4861 and [O\,\textsc{iii}]$\lambda$5007, especially in ESO\,157-49, ESO\,443-21, ESO\,469-15, IC\,217, IC\,1553 and PGC\,30591.
This resulted in many poorly fitted Voronoi bins.
We then tested different setups for \textsc{gandalf}:
\begin{itemize}
    \item Fit kinematics of all lines free
    \item Fit kinematics of [O\,\textsc{iii}]$\lambda$$\lambda$4959,5007  \& [N\,\textsc{i}]$\lambda$5198 tied to H$\beta$ 
    \item Fit kinematics of [N\,\textsc{i}]$\lambda$5198 \& [O\,\textsc{iii}]$\lambda$4959 tied to [O\,\textsc{iii}]$\lambda$5007
    \item Fit kinematics of [O\,\textsc{iii}]$\lambda$4959 tied to [O\,\textsc{iii}]$\lambda$5007 \& fit [N\,\textsc{i}]$\lambda$5198 free
\end{itemize}
But none of them improved the fitting of the emission lines.\\

Further, we tested to fit the emission lines with \textsc{ppxf} \citep{cappellari_2017, cappellari_2022} using one gas component.
Using the variable LSF (Line Spread Function) of MUSE, the stellar continuum was fitted with combinations of SSP models, while Gaussian templates are added to this continuum by \textsc{ppxf} to fit the emission lines. 
For each of the 4 emission lines H$\beta$ $\lambda$4861, [O\,\textsc{iii}]$\lambda$$\lambda$4959,5007 and [N\,\textsc{i}]$\lambda$5198, a separate Gaussian template was created and fitted. 
Those templates were fixed to the same kinematics but allowed to vary in amplitude. 
We obtained the flux, velocity $V$, and velocity dispersion $\sigma$ of the ionized gas.
Using \textsc{ppxf}, the galaxies Voronoi bin spectra were fitted better than before.
Still, ESO\,157-49, ESO\,469-15, IC\,217, ESO\,443-21 and PGC\,30591 showed large residual in the regions of [O\,\textsc{iii}]$\lambda$$\lambda$4959,5007 emission.
Further, for IC\,1553, all the emission lines were fitted poorly for many Voronoi bins, since the lines were asymmetric and had too much deviation from a true Gaussian shape, such that only one Gaussian template did not fit the lines well enough.
Because of that, we tested for this galaxy the use of two gas components, i.e., two Gaussian templates with different kinematics, per emission line.
However, this approach still gave large residuals of the fits.
Only ESO\,544-27 and PGC\,28308 showed good fits without any larger residuals for all emission lines.

After fitting the gas emission lines, the ``gas best-fit'' spectra (consisting of the best-fitting Gaussian gas templates) were subtracted from the observed galaxy spectra for each Voronoi bin, and emission-cleaned spectra were obtained for all galaxies.
In general, the gas emission was quite high for some galaxies and had fluxes up to 20 times higher than the stellar continuum.
Because of this, the subtraction of the ``gas best-fit'' spectra left large wiggles in the emission cleaned spectrum, even if the gas emission lines themselves showed good fits (similar to what was reported by \citealt{gomez_2023}).
This is because the residuals of the gas emission were larger than the noise of the spectra.
For this reason the emission-cleaned stellar spectra of ESO\,157-49, ESO\,443-21, ESO\,469-15, IC\,217, IC\,1553 and PGC\,30591 showed large residual wiggles, especially in the [O\,\textsc{iii}]$\lambda$$\lambda$4959,5007 emission regions, and for IC\,1553 also the H$\beta$~$\lambda$4861 emission region.
Our strategy to deal with these wiggles will be explained in Sect.~\ref{subsec:ppxf_pop}.

\subsubsection{Extinction correction}
\label{subsec:ppxf_extinction}
Because late-type galaxies usually contain large amounts of gas and dust, we corrected the galaxy spectra for interstellar extinction before fitting for stellar populations.
Therefore, we followed the same procedure used in \citet{sattler_2023}, where we first fit each Voronoi bin spectrum for extinction with \textsc{ppxf} by using the built-in reddening parameter.
We then removed the extinction from the spectra using version 0.4.6 of the \textsc{extinction}\footnote{\url{https://extinction.readthedocs.io/en/latest/}} package.
Following \citet{emsellem_2022}, we also corrected all spectra for Galactic extinction of the Milky Way by division with the \citet{cardelli_1989} extinction law.

\subsection{Stellar population analysis}
\label{sub:sp_analys}

\subsubsection{Stellar population fitting}
\label{subsec:ppxf_pop}
To extract the stellar populations (age, metallicity, and [Mg/Fe] abundance) from the extinction-corrected spectra, we fit those with \textsc{ppxf} including a regularization \citep{cappellari_2017}.
This regularization leads to smoother weight distribution on the different SSPs and thus smoother stellar population solutions.
Regularization helps to avoid spurious stellar population solutions that are unstable against changes in the fitting process.
A good choice of the regularization parameter is necessary to avoid smoothing to the extent that minor stellar population components disappear in the distributions of age, metallicity, or [Mg/Fe] abundance. 
To find the maximum value of regularization, we followed the approach given in the \textsc{ppxf} code itself \citep{cappellari_2017, cappellari_2022}, also used by \citet{pinna_2019b, pinna_2019a}, \citet{boecker_2019}, and \citet{sattler_2023}. We applied this approach to one central Voronoi bin of each galaxy, following the steps below: 
\begin{enumerate}
    \item A fit was performed with noise = 1 and regularization = 0, so that $\chi^2$ = $N_{\text{goodpix}}$
    \item The noise was re-scaled to $\sqrt{\frac{\chi^2}{N_{\text{goodpix}}}}$, while the \textsc{ppxf} fit was iteratively repeated increasing regularization, until 
    $\Delta \chi^2 \approx  \sqrt{2N_{\text{goodpix}}}$
    
\label{eq:regul}
\end{enumerate}

where $\chi^2$ is a measure of the goodness of the fit and $N_{\text{goodpix}}$ is the number of pixels which are included in the fit.
This final value of regularization was the maximum regularization parameter.
A variety of values lower than the upper limit still give results that are consistent with the data.
After testing different values of regularization (see App.~\ref{appendix:pop}), we decided to use a value of 1 for all galaxies, as it applied some smoothing but allowed the identification of features in the star formation histories.
\\\\
To recover those spatial bins with large residuals resulting from the emission line subtraction, we masked in the emission-cleaned spectra of ESO\,157-49, ESO\,443-21, ESO\,469-15, IC\,217 and PGC\,30591 the regions of [O\,\textsc{iii}]$\lambda$$\lambda$4959,5007 emission with a window of 400~km~s$^{-1}$ for the stellar population fitting.
For IC\,1553, we used the original Voronoi binned spectra for the fitting and masked the region of H$\beta$ $\lambda$4861, [O\,\textsc{iii}]$\lambda$$\lambda$4959,5007 and [N\,\textsc{i}]$\lambda$5198 emission.
A summary of the setup can be seen in Tab.~\ref{tab:settings}.

We fit all galaxies for light- and mass-weighted stellar population parameters. 
The chosen SSP models (Sect.~\ref{subsec:models}) are normalized to have 1~$M_{\odot}$ at birth.
So, any further normalization was not necessary to extract mass-weighted stellar population parameters. 
We normalized the SSP models by their mean flux to obtain light-weighted results. 
Also, we applied a multiplicative polynomial of 8$^{\text{th}}$ degree and no additive polynomials during the fitting process and discarded all Voronoi bins that were ill-fit (the shape of the spectrum was not followed or large residuals were present).

\begin{table}
    \caption{Summary of the settings for the sample. }
    \begin{tabular}{p{1.8cm} p{1.1cm} p{1.3cm} p{2.5cm}}
    \hline
    \\
    \hfill{} Galaxy & S/N $^{\left(1\right)}$ & \#\,bins $^{\left(2\right)}$ & masked for pop $^{\left(3\right)}$\\
    \hline
    \\
    \hfill{} ESO\,157-49 & 60 & 1422 & [O\,\textsc{iii}]\\
    \hfill{} ESO\,469-15 & 60 & 559 & [O\,\textsc{iii}]\\
    \hfill{} ESO\,544-27 & 60 & 476 & -\\
    \hfill{} IC\,217 & 60 & 252 & [O\,\textsc{iii}]\\
    \hfill{} IC\,1553 & 100 & 387 & H$\beta$ + [O\,\textsc{iii}] + [N\,\textsc{i}] \\
    \hfill{} PGC\,28308 & 60 & 188 & -\\
    \hfill{} ESO\,443-21 & 60 & 268 & [O\,\textsc{iii}]\\
    \hfill{} PGC\,30591 & 60 & 176 & [O\,\textsc{iii}]\\
    \hline
    \\
    \end{tabular}
    \begin{tablenotes}
            \item \textbf{Notes:} (1) Target-S/R for the Voronoi bins. (2) Number of Voronoi bins. (3) Emission line regions that were masked for the stellar population fitting.
    \end{tablenotes}
    \label{tab:settings}
\end{table}

\begin{figure*}
    \centering
    \includegraphics[width=0.85\textwidth]{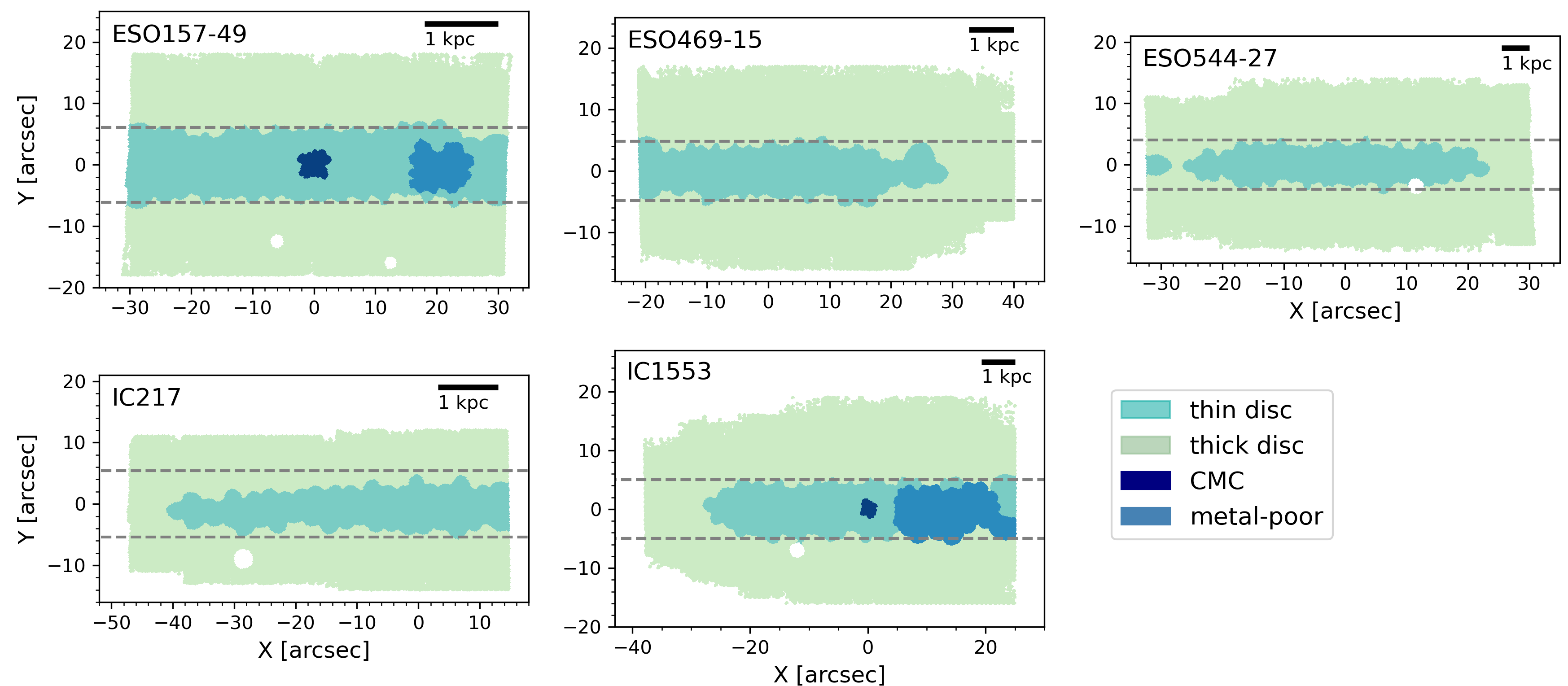}
    \caption{Component decomposition for the five out of the eight sample galaxies that contain distinct thick and thin disks.}
    \label{fig:components}
\end{figure*}

\subsubsection{Star formation histories}
\label{subsec:SFH}
For the extraction of the star formation histories, normalized by the mass of the covered region of the galaxy, we first calculated the mass-to-light ratio for each Voronoi bin by combining the mass-to-light ratios of the MILES SSP models \citep{vazdekis_2015} with the mass-weighted stellar population results.
To calculate the mass of each Voronoi bin as in \citet{pinna_2019a} and \citet{sattler_2023}, we additionally used \textit{g}-band images (see Fig.~\ref{fig:g_band}) from Pan-STARRS1\footnote{\url{https://outerspace.stsci.edu/display/PANSTARRS/Pan-STARRS1+data+archive+home+page}} \citep{panstarrs_2016, panstarrs_2020a, panstarrs_2020b, panstarrs_2020c, panstarrs_2020d, panstarrs_2020e} for ESO\,443-21, ESO\,469-15, ESO\,544-27, IC\,217, IC\,1553, PGC\,28308 and PGC\,30591.
For ESO\,157-49, not part of the Pan-STARRS1 sample, we used a \textit{g}-band image from the Dark Energy Survey DR2\footnote{\url{https://www.darkenergysurvey.org/}} \citep{DES_2018, DES_2021}.
Then, we converted the \textit{g}-band magnitudes into \textit{V}-band magnitudes, following \citet{pinna_2019a}.
Taking into account the distances to the galaxies, we calculated the total mass in each MUSE pixel.
These masses per pixel were then multiplied by the number of pixels for each Voronoi bin to get the mass of the entire Voronoi bin.
\\\\
To obtain the star formation histories of the different disk components, we defined the regions of the thin and thick disks using the heights above and below which the thick disk dominates $z_{\text{c1}}$ (see Tab.~\ref{tab:prop}).
For galaxies with CMCs (see Sect.~\ref{sec:sample}), although these CMCs do not dominate at any time the vertical surface brightness profiles, we decided to ignore the central region of the thin disks, to avoid any potential contamination from the CMCs. 
Therefore, we discarded all Voronoi bins in a central region within $r_{\text{CMC}}$, the scale radius of the S\'ersic profile of the CMC \citep{comeron_2018}. 
However, as the Voronoi bins are already quite large in the central regions, this CMC region is not a perfect circular cut.
Since the thin disks of ESO\,157-49 and IC\,1553 showed in the metallicity maps (presented in Sect.~\ref{subsec:res_pop}) a region with a much lower metallicity than other comparable regions in the thin disk, we found it interesting to also extract the star formation histories of these metal-poor regions.
The different regions for which we traced the star formation histories are mapped in Fig.~\ref{fig:components}. 
To obtain the star formation histories, we summed up the weights from the mass-weighted results for each Voronoi bin belonging to the component, so all the mass fractions corresponding to the same age bin with different metallicities and [Mg/Fe] abundances.
Mass-weighted maps are shown in Fig.~\ref{fig:mw_maps}. 
It is important to note that the values for each Voronoi bin in the maps are the weighted averages of age, metallicity, and [Mg/Fe], resulting from the weighted combination of all SSP models. 
On the other hand, the star formation histories correspond to the SSP distributions across ages. 
Thus, an age distribution that only contains very young and very old stars, for instance, can lead to an intermediate age Voronoi bin in the age map. 
Then, we computed a mass-weighted average of the mass fractions per age for all Voronoi bins normalized by the galaxy mass. 
As the age resolution changes from 0.01~Gyr up to 0.5~Gyr in the latest 4~Gyr (after an age of 4~Gyr, the resolution stays consistent with 0.5~Gyr), we further summed up the mass fractions of the different age bins to achieve a stable resolution among all ages.
This will help to better compare the mass fractions for all the ages with different resolutions, and is also a key factor for a proper extraction of the star formation history \citep{Wang2024}.

\subsubsection{Monte Carlo simulations}
\label{subsec:MC_simulations}
To estimate the uncertainties in the stellar populations and star formation histories, we performed Monte Carlo simulations.
Hence, we followed these steps for each Voronoi bin of the galaxies for 1000 realizations:
\begin{enumerate}
    \item From the stellar population fits with \textsc{ppxf}, calculate the wavelength-dependent residuals as the difference between the best-fit model spectrum and the observed galaxy spectrum
    \item For each realization, calculate the noise spectrum as a random noise from a Gaussian distribution with a standard deviation equal to the residuals at each wavelength. 
    \item Add this random noise to the observed spectrum, so that a different noise is applied for each realization. 
    \item Fit this new noisy spectrum with \textsc{ppxf} for stellar populations using no regularization.
\end{enumerate}
Thereafter, we calculated the uncertainties in age, metallicity, and [Mg/Fe] abundance for each Voronoi bin as the standard deviation of the age, metallicity, and [Mg/Fe] distributions from these 1000 realizations.
To obtain the uncertainties of the star formation histories, we computed a separate star formation history for each of the 1000 Monte Carlo realizations and determined the standard deviation from the mean mass fractions for each age.

\section{Results}
\label{sec:results}

\begin{table}
    \centering
    \caption{Average uncertainties for the light-weighted stellar population maps (Fig.~\ref{fig:lw_maps}) calculated as the mean uncertainty of all bins belonging to the corresponding component. The detailed uncertainty maps can be seen in App.~\ref{appendix:uncertainties}.}
    \begin{tabular}{p{1.8cm} p{1.6cm} p{0.8cm} p{1cm} p{1.3cm}}
        \hline
		\\
		\hfill{} Galaxy & Component & $\Delta$Age & $\Delta$[M/H] & $\Delta$[Mg/Fe]\\
        \hfill{} \ & \ & [Gyr] & [dex] & [dex]\\
        \hline
        \\
		\hfill{} ESO\,157-49 & thin disk \newline thick disk & 0.30 \newline 1.23 & 0.05 \newline 0.10 & 0.12 \newline 0.13\\
		\hfill{} ESO\,469-15  & thin disk \newline thick disk & 0.33 \newline 1.85 & 0.05 \newline 0.17 & 0.12 \newline 0.19 \\
        \hfill{} ESO\,544-27  & thin disk \newline thick disk & 0.31 \newline 1.52 & 0.05 \newline 0.11 & 0.07 \newline 0.18 \\
        \hfill{} IC\,217  & thin disk \newline thick disk & 0.30 \newline 0.93 & 0.07 \newline 0.08 & 0.10 \newline 0.13 \\
        \hfill{} IC\,1553  & thin disk \newline thick disk & 0.93 \newline 3.02 & 0.24 \newline 0.48 & 0.31 \newline 0.23 \\
        \hfill{} PGC\,28308  & full galaxy & 0.32 & 0.04 & 0.14 \\
        \hfill{} ESO\,443-21 & full galaxy & 0.48 & 0.07 & 0.24 \\
        \hfill{} PGC\,30591 & full galaxy & 0.30 & 0.05 & 0.08 \\
        \hline
		\\
    \end{tabular}
    \label{tab:uncertainties_lw}
\end{table}

\begin{figure*}
    \centering
    \includegraphics[width=0.8\textwidth]{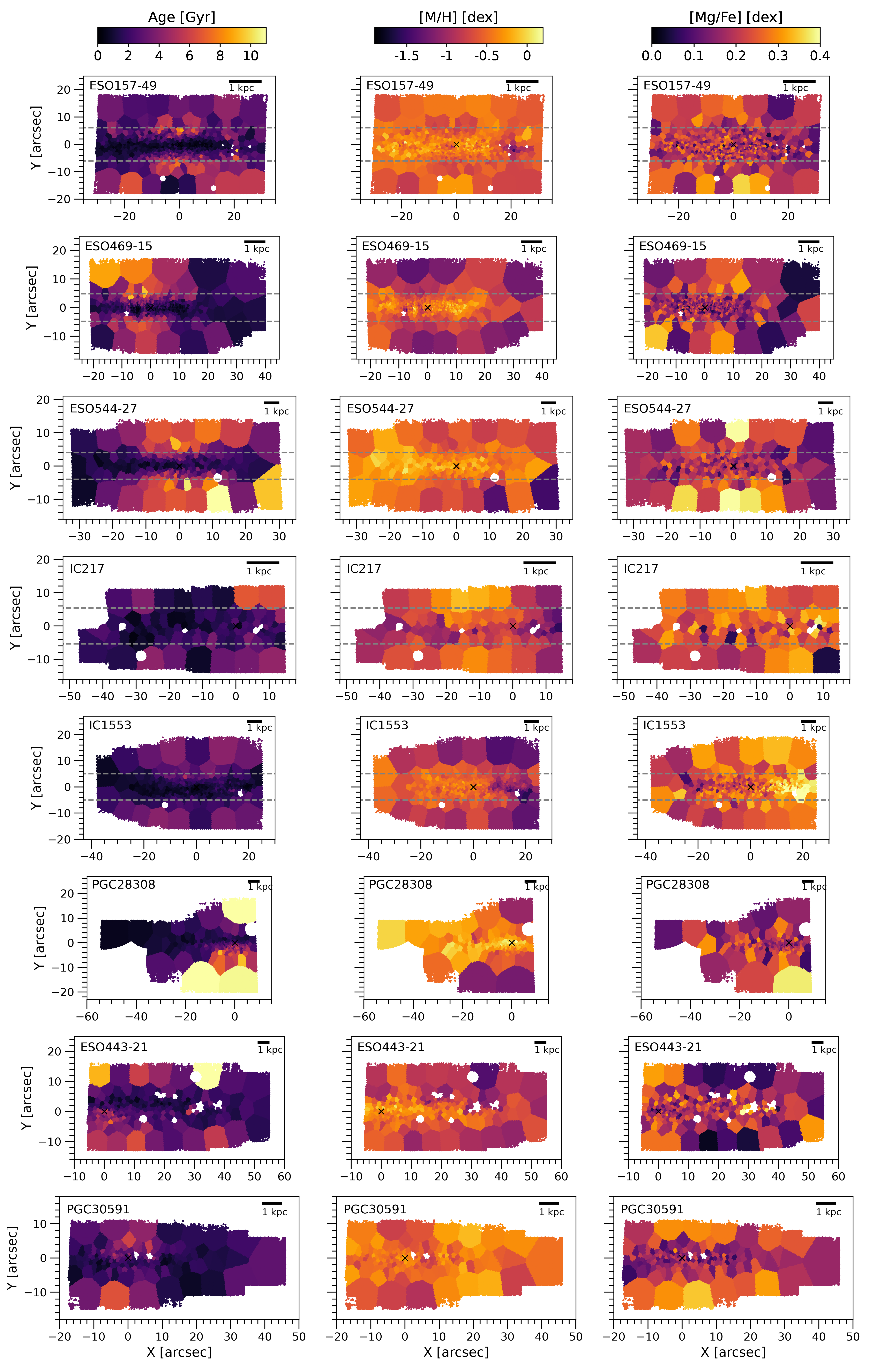}
    \caption{Light-weighted age (left), metallicity (middle), and [Mg/Fe] abundance (right) maps for the full sample. For the galaxies with distinct morphological thick and thin disks, the dashed gray lines mark the regions above and below which the thick disk dominates the vertical surface brightness profiles.}
    \label{fig:lw_maps}
\end{figure*}

\subsection{Mapping stellar populations}
\label{subsec:res_pop}
In this section, we present the light-weighted stellar population maps (Fig.~\ref{fig:lw_maps}).
Mass-weighted results, which were used for the star formation histories, are shown in App.~\ref{appendix:mw}.
Average uncertainties for the light-weighted stellar population maps can be seen in Tab.~\ref{tab:uncertainties_lw}, and maps of the uncertainties for each Voronoi bin can be found in Fig.~\ref{fig:MC_lw_maps}.
As the uncertainties in the stellar populations can be large, especially for IC\,1553, we need to be cautious with quantitative analysis based on the absolute values and set the focus on the relative differences between the thin and thick disks. 
\\\\
Taking a look at the light-weighted age maps (Fig.~\ref{fig:lw_maps}, left column), most galaxies with clear distinct thin and thick disks (ESO\,157-49, ESO\,469-15, ESO\,544-27, IC\,1553) show on average very young stars (<~2~Gyr) in the thin disk and older stars (in a wide range from  3~to~11~Gyr) in the region of the thick disk.
Thereby, the thick disk of IC\,1553 shows the youngest ages staying below 4~Gyr.
IC\,217, which also has two disk components in the vertical surface brightness profiles, shows no clear differences in thin and thick disk ages; the whole galaxy is rather young, with ages below 7~Gyr.
For ESO\,157-49, the oldest stars with ages around 6~to~8~Gyr are located in the inner transition region between the thick and the thin disk, and might be related to a central component (Sect.~\ref{sub:SB_profiles}).
The boxy shape suggests this might be a boxy bulge. 
In the thick disk of ESO\,544-27, the ages decrease from around 8~Gyr at smaller radii to below 3~Gyr at larger radii, suggesting a thin disk flaring. 
For IC\,1553, a similar behavior is observed in the region of the thick disk only above the thin disk.
Ages decrease toward larger radii (from about 4~Gyr to very recent ages), suggesting a warp.\\ 
Galaxies without clear distinct thin and thick disks (PGC\,28308, ESO\,443-21 and PGC\,30591) show in general, a similar structure as the rest of the sample, with younger stars along the midplane and older stars in regions at larger heights.
However, there is some more variance from galaxy to galaxy. 
For PGC\,28308 in particular, the oldest stars (as old as 11~Gyr) are located at the largest distances from the midplane in a spheroidal shape.
\\\\
In the light-weighted metallicity maps (Fig.~\ref{fig:lw_maps}, middle column), most galaxies with clear distinct thin and thick disks (ESO\,157-49, ESO\,469-15, ESO\,544-27 and IC\,1553) show higher metallicities (nearly solar) in the thin disks and lower sub-solar metallicities in the thick disks.
Besides, ESO\,157-49 and IC\,1553 show a region with very low metallicity (even lower than the average of the thick disk) in their thin disks around 20~arcsec (for ESO\,157-49) and 5-25~arcsec (for IC\,1553) to the right from the galaxy center.
In contrast to the other galaxies with distinct thin and thick disks, interestingly, IC\,217's thin disk is slightly metal-poorer than the thick disk.
Having no distinct thin and thick disk components, the galaxies PGC\,28308 and ESO\,443-21 follow the same trend with higher (nearly solar) metallicities in the central midplane regions and lower clearly sub-solar metallicities for regions at larger heights.
For PGC\,30591, there are no clear metallicity differences with sub-solar values all over the disk.
\\\\
The light-weighted [Mg/Fe] maps are shown in the right column of Fig.~\ref{fig:lw_maps}.
ESO\,157-49, ESO\,469-15, ESO\,544-27, IC\,1553, and PGC\,30591 show in general lower [Mg/Fe] abundances in the thin disks or midplane regions and higher abundances for regions at larger heights.
In IC\,1553, the most [Mg/Fe]-rich region is located in the thin disk and spatially matches the most metal-poor region that we have mentioned in the previous paragraph. 
However, all the [Mg/Fe] distributions are quite noisy and due to the uncertainties being larger than 0.1 dex for most of the galaxies, part of these differences are not significant.
For the remaining galaxies (IC\,217, PGC\,28308, and ESO\,443-21), no signs of structures in the [Mg/Fe] distributions can be identified.

\subsection{Star formation histories}
\label{subsec:res_SFH}

\begin{figure*}
    \centering
    \vspace{-0.5cm}
    \includegraphics[width=1\textwidth]{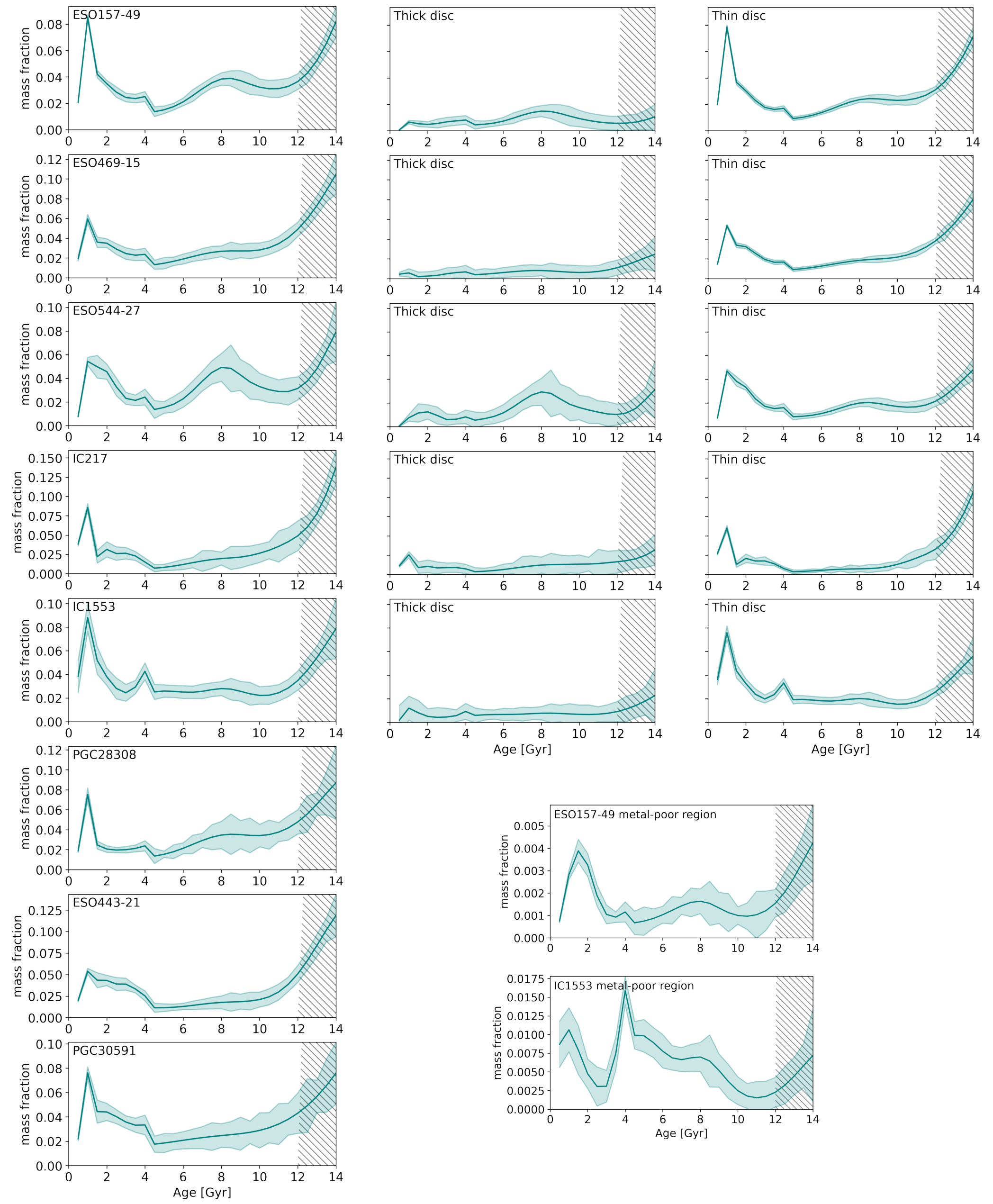}
    \caption{Star formation histories in terms of mass fraction per age with evenly spaced age bins every 0.5~Gyr. The uncertainties from the Monte Carlo simulations are shown by the shaded light blue areas.
    The left column shows the star formation histories for the fully covered regions of the galaxies, while the middle and right columns show the thick and thin disks, respectively, for the five out of the eight galaxies containing two well-defined disk components (they share the same y-axis with the left column). The additional separate panels on the lower right corner show the star formation histories of the metal-poor regions for ESO\,157-49 and IC\,1553.
    Large mass fractions at old ages should be handled with caution, as there might be a bias of \textsc{ppxf} fitting toward very old ages (see Sect.~\ref{subsec:res_SFH}) and are shaded with a gray area.}
    \label{fig:SFH_maps}
\end{figure*}

The star formation histories for the whole stellar disks and the different components as defined in Sect.~\ref{subsec:SFH} and Fig.~\ref{fig:components} (thin disk, thick disk, and metal-poor region) of the sample galaxies are presented in Fig.~\ref{fig:SFH_maps}. 
In these plots, the mass fractions for each of the components are shown as functions of age (with a stable resolution of 0.5 Gyr as mentioned in Sect.\ref{subsec:SFH}) together with the uncertainties as shaded regions.
It is important to recall that these star formation histories were extracted from the mass-weighted results (App.~\ref{appendix:mw}), where average ages are older than the ones shown in Fig.~\ref{fig:lw_maps}. 
Also, as already mentioned in Sect.~\ref{subsec:SFH}, the star formation histories are a recovery of the whole SSP distribution.
\\
Further, it should be pointed out that we also analyzed the chemical evolution from the distributions of metallicity and [Mg/Fe] abundances associated with different ages, following \citet{sattler_2023}.
However, metallicity and [Mg/Fe] abundances for different ages had large uncertainties, and we decided to show only the mass fractions here. 
This might be due to the quality of the data, the presence of a large mass fraction of young stars for which chemical properties are more difficult to determine (see also discussion in \citealt{sattler_2023}), combined with age-metallicity degeneracy (this degeneracy is discussed in App.~\ref{appendix:templates}).
The fractional contributions of thin and thick disks and metal-poor regions to the stellar mass of the analyzed region of the galaxy are contained in Tab.~\ref{tab:masses}.
There, the mass fractions given for the combined thin disk and thick disk of ESO\,157-49 and IC\,1553 are smaller than 100~\%, because the central regions including the CMCs were excluded from the thin disk component.
The metal-poor regions' mass is, however, included in the mass fraction of the thin disk.
\\\\

\begin{table}
    \centering
    \caption{Mass fractions of the different regions.}
    \begin{tabular}{p{1.8cm} p{1.5cm} p{1.5cm} p{1.7cm}}
    \hline
    \\
    \hfill{} Galaxy & thick disk & thin disk & metal-poor\\
    \hline
    \\
    \hfill{} ESO\,157-49 & 0.22 & 0.76 & 0.05\\
    \hfill{} ESO\,469-15 & 0.23 & 0.77 & -\\
    \hfill{} ESO\,544-27 & 0.39 & 0.61 & -\\
    \hfill{} IC\,217 & 0.33 & 0.67 & -\\
    \hfill{} IC\,1553 & 0.22 & 0.77 & 0.19\\
    \hline
    \\
    \end{tabular}
    \label{tab:masses}
\end{table}
In the star formation histories of the full galaxies (Fig.~\ref{fig:SFH_maps}, left column), star formation started at very early times, with a very large amount of stars formed at old ages for all galaxies.
After that, the star formation dropped around 11~Gyr and might have increased again at intermediate ages (around 8-9~Gyr ago) for ESO157-49, ESO\,544-27, and IC\,1553.
For IC\,1553, this higher star formation forms a plateau emerging between around 4 to 9~Gyr.
However, due to the uncertainties (shaded areas), the star formation could rather be more constant along these intermediate ages.
On the other hand, and even with the uncertainties in mind, ESO\,469-15, IC\,217, PGC\,28308, ESO\,443-21, and PGC\,30591 have a flatter slope with roughly constant star formation at intermediate ages.
At the youngest ages, all galaxies show a final burst at more recent times. 
\\

If we look at the star formation histories of the thick disks (Fig.~\ref{fig:SFH_maps}, middle column), fewer stars are formed at the oldest ages compared to the whole galaxy.
At intermediate ages (around 8~Gyr, as a reference, but see the discussion below, about uncertainties), the thick disks of ESO\,157-49, and ESO\,544-27 show an increase in star formation.
Most of the mass in the thick disk might have formed during this star-forming episode in ESO\,157-49 and ESO\,544-27.
Taking into account the uncertainties, these galaxies could have also formed their thick disk stars similar to ESO\,469-15, IC\,217, and IC\,1553, which show a relatively steady thick disk star formation history without any rapid increase along intermediate ages.
Afterward, ESO\,157-49 and ESO\,469-15 show a decrease in star formation with nearly no stars formed at recent times.
The thick disks of ESO\,544-27, IC\,217, and IC\,1553 display a late star-forming episode at recent times. 
However, mass fractions have large uncertainties for these young ages (shaded regions).\\
In the thin disks (Fig.~\ref{fig:SFH_maps}, right column), mass weights are in general distributed mostly in the very old and young populations. 
However, it should be taken into account that the separation of thin and thick disks is not a straightforward process, and the thin disk spectra contain contamination of thick disk light.
For ESO\,157-49 and ESO\,544-27, the increase in star formation around 8~Gyr in the thin disks is not as pronounced as in the thick disks, while for IC\,1553 we see a prominent plateau along intermediate ages. 
The thin disks of ESO\,469-15 and IC\,217 show a continuous decrease in star formation from the oldest ages to around 4~Gyr.
Thereby, IC\,217 seems to get almost totally quenched along intermediate ages.
At the youngest ages, all galaxies show a final starburst, with a different time scale and strength for different galaxies. 
Although some stars were also formed at thick disk heights during these final starbursts, it is much more pronounced in the thin disks compared to the thick disks.\\
For two galaxies, we extracted the star formation history of the metal-poor regions in the thin disk (Sect.~\ref{subsec:SFH}).
These are shown in the lower right corner of Fig.~\ref{fig:SFH_maps}.
In ESO\,157-49, the final starburst at the youngest ages contributed more stellar mass compared to intermediate ages.
For the metal-poor component in IC\,1553, two separate peaks might emerge between around 4 to 9~Gyr, showing the highest amount of star formation compared to other ages.
\\\\
For nearly all galaxies (except ESO\,157-49 and IC\,1553), the highest peak in the star formation histories of the full galaxies and thin disk is located at the oldest ages, suggesting very intense star formation at the beginning of galaxy evolution. 
However, this large mass fraction at the oldest ages might be due to a well-known and tested bias from \textsc{ppxf} toward very old stellar populations \citep{pinna_2019a, Wang2024}.
Especially in \citet{Wang2024}, it was shown that the age distribution is overestimated in the ranges between 2 to 4~Gyr and 12 to 14~Gyr when using \textsc{ppxf} and MILES SSP models.\\ 
Uncertainties in the mass fractions are quite large for the thick disks ($\pm$~1~\% of the galaxy mass) and especially for the oldest ages (up to $\pm$~2~\% of the galaxy mass). 
For the thin disks, these are much lower on average ($\pm$~0.2~\% of the galaxy mass) and similar to the ones for the full galaxies, while the uncertainties for the metal-poor regions vary between $\pm$~0.1~\% and $\pm$~0.2~\% of the galaxy mass. 
However, the shades in Fig.~\ref{fig:SFH_maps} also suggest that the uncertainties in the position of peaks in the star formation are quite large for some of them (about 3-4~Gyr, where shades are thicker) and small for others (where shades are thin). \\
Additionally, we have performed detailed tests on mock spectra with different initial star formation history slopes (see App.~\ref{appendix:SFHmock}). 
These show that peaks in star formation are, in general, recovered, but in some cases with large age error bars of up to 3-4~Gyr in their position (similar to uncertainties estimated via Monte Carlo simulations for peaks with thick shades in Fig.~\ref{fig:SFH_maps}). 
On the other hand, metallicity is well recovered in general. 
Having this in mind, a quantitative analysis of the results of the star formation histories needs to be handled with caution, as there might be some limitations with the fitting method. 
However, a qualitative discussion such as the one presented here provides a nice overview of the formation of galaxies and galaxy components. 
Future improvements, such as higher S/N and particularly a wider wavelength coverage, could help to enhance the accuracy of star formation history recoveries and reduce some of these limitations.

\section{Discussion}
\label{sec:discussion}

\subsection{Thin and thick disks in comparison with other edge-on galaxies}
This is the first systematic study with integral-field spectroscopy of thick and thin disks in very late-type galaxies. 
We analyzed a sample of eight edge-on galaxies, of which five have distinct thick and thin disks. 
Our results showed thin disks that are mostly younger, more metal-rich (except for IC\,217) and less enhanced in their [Mg/Fe] abundance than thick disks.
This behavior was found in most observed thin and thick disks (see Sect.~\ref{sec:introduction}). 
However, compared to previous studies on earlier-type galaxies, mean ages are shifted toward young stars, and differences between thin and thick disks in the stellar population parameters are lower in our sample. 
While lenticular galaxies were found to have a very old (older than 10~Gyr), metal-poor and $\alpha$-enhanced thick disk and a metal-rich thin disk that can be significantly younger (but not necessarily, \citealt{comeron_2016, pinna_2019b, pinna_2019a}), there are very few integral-field spectroscopy studies of stellar populations in edge-on spiral galaxies published to date, covering the region dominated by the thick disk. 
As spiral galaxies, both NGC\,5746 \citep{martig_2021} and ESO\,533-4 \citep{comeron_2015} showed an old thick disk (10~Gyr or older) and a much younger thin disk. 
NGC\,3501, a late-type spiral galaxy with no morphological distinction between a thick and a thin disk in the observed region, displayed interesting features different from what was previously found. 
These were overall young ages (<~8~Gyr for most of the Voronoi bins) and low metallicities, with a small, thinner disk forming in the inner region of the midplane \citep{sattler_2023}. 
\\
The galaxies in our sample are somehow different from previously studied edge-on galaxies, since they are gas-rich, very late-type galaxies, with very strong emission lines and extraplanar ionized gas covering the region of the thick \citep{comeron_2019, rautio_2022}. 
Our stellar population results show that these galaxies are systematically younger than most edge-on galaxies previously observed with integral-field spectroscopy. 
The thin disks in our study show Voronoi bins with mean ages younger than 2~Gyr.
Also, the thick disks show relatively young aged Voronoi bins (on average $\sim$~4~Gyr in Fig.~\ref{fig:lw_maps} with only a few Voronoi bins reaching a mean age of $\geq$~8~Gyr), which are interestingly similar to thin disks in other galaxies \citep[e.g.,][]{pinna_2019b, pinna_2019a, martig_2021}.
Thick disks with these young ages in spiral galaxies, with a small age difference to the thin disks, were not mapped before with integral-field spectroscopy, while some long-slit studies found hints of such thick disk ages in late-type spirals \citep[e.g.,][]{yoachim_2008b}. 
If we compare thick disk ages measured in this work to the ones of more massive galaxies, they seem to qualitatively agree with the predictions made for late-type galaxies in \citet{comeron_2021}, where thick disk ages are younger for lower-mass galaxies.\\
Differences in metallicity and [Mg/Fe] abundance between thick and thin disks are not sharp in our sample.
These galaxies are overall metal-poor and only their thin disks have nearly solar metallicities, which is lower than the thin disks in earlier-type galaxies in prior studies (super-solar in \citealt{pinna_2019b,pinna_2019a,martig_2021}). 
Low metallicities are characteristic of very late-type galaxies and their thick and thin disks \citep{yoachim_2008b, comeron_2015}. 
Moreover, our thick disks are also not as [Mg/Fe]-enhanced as the four earlier-type galaxies for which spatially resolved [Mg/Fe] was measured \citep{pinna_2019b,pinna_2019a,martig_2021}.

\subsection{Formation and evolution of thick and thin disks}
\label{subsec:discuss_formation}

The combination of young ages (<~8~Gyr) with low metallicities (<~0~dex) and low [Mg/Fe] abundances (<~0.3~dex), for the majority of Voronoi bins in the thick disks, suggests a slow evolution extended in time.
As stellar masses of our sample galaxies are quite low when compared to other late-type galaxies \citep{comeron_2021}, this slow evolution was also supported by \citet{Padmanabhan1993}, stating that the mass assembly processes of galaxies are delayed in regions of low mass density.\\
Star formation histories of ESO\,157-49 and ESO\,544-27 
show that these galaxies might have had an intense star formation episode in their thick disk, as a reference, around 8~Gyr ago (with large uncertainties as discussed in Sect.~\ref{subsec:res_SFH}).
This also happened in the metal-poor regions of ESO\,157-49 and IC\,1553.
While this peak is not as strong for ESO\,157-49 as in IC\,1553, and could also indicate rather constant star formation, the latter experienced its strongest star formation peaks extended over a longer period, between around 4 and 8~Gyr ago.
ESO\,469-15, IC\,217, and IC\,1553, rather show a smoother star formation history in their thick disks, with a small stellar component around 6-8~Gyr. 
Thus, we found stellar populations formed around 7-11~Gyr ago in all thick disks in our sample, although more or less significant depending on the galaxy.
There might have been an event around 8~Gyr ago that potentially contributed to the formation of the thick disk and the prominent two disk structure. 
Some possible mechanisms could be mergers with satellites \citep{brook_2004, quinn_1993, abadi_2003} or gas accretion from companions (e.g., for ESO\,157-49, see below) or the intergalactic medium.\\
Besides, our tests in App.~\ref{appendix:SFHthinheight} show that these characteristics of the thin and thick disk star formation histories are robust against variations in the geometric definition of thin disks. 
However, for all star formation histories it is important to keep in mind the presence of systematics and limitations in the fitting method (see App.~\ref{appendix:SFHmock}).\\
The stellar kinematics of our sample, presented in \citet{comeron_2019}, did not show any evidence that direct star accretion from satellite galaxies is the main driver of thick disk formation.
No galaxy shows signs of a retrograde stellar component larger than 10~\%, but this does not rule out the accretion of a small fraction of stars. 
Also, a small retrograde component does not exclude the formation of thick disks from interactions or mergers with other stellar systems, as prograde mergers are favored due to dynamical friction \citep{comeron_2019}.
As mentioned in Sect.~\ref{sec:introduction} and different from our sample, the thick disks in \citet{pinna_2019b, pinna_2019a} and \citet{ martig_2021} accreted around 30~\% of their mass during a merger event.
However, we have shown here that the thick disks in our sample are different from those of S0 galaxies in clusters \citep{pinna_2019b, pinna_2019a}. 
They show much younger mean ages and extended star formation histories, hinting at a dominant contribution from stars formed in-situ. 
In this sample of late-type spiral galaxies, which are overall young and gas-rich, different star formation peaks might have been triggered or fueled by different episodes of gas accretion.

Star formation at the youngest ages (younger than 4~Gyr) is present in all galaxies, even in the regions of the thick disks, and significantly more enhanced in their thin disks.
Thereby, IC\,1553 has the highest SFR~=~0.96 $M_{\odot}$/yr among the galaxies with distinct thin and thick disks (see Tab.~\ref{tab:prop}, \citealt{rautio_2022}).
Star formation in the last few billion years was also found by \citeauthor{zaragoza-cardiel_2020} (\citeyear{zaragoza-cardiel_2020}, see also Sect.~\ref{sec:sample}) for IC\,217 and IC\,1553. 
Moreover, \citet{rautio_2022} observed vertical rotation lags of the ionized gas for ESO\,157-49, ESO\,544-27, IC\,217, and IC\,1553, and proposed gas accretion to explain them. 
\citet{keeney_2013} and \citet{lopez_2020} found signs of a galactic fountain and outflows for ESO\,157-49. 
Therefore, gas accretion might be responsible for recent star formation events, which are primarily present in the thin disks. 
\\
Putting all of this together, we propose that most thick disk stars formed in the same galaxies during a phase of intense star formation that might have peaked around old and intermediate ages (around 7-11~Gyr ago). 
This scenario was already proposed by \citet{brook_2004} and multiple simulations (see Sect.~\ref{subsec:numerical_sim}), showing that galaxy formation happens in two phases, beginning with a period of intense turbulent star formation.
During this early bursty phase, the orbits of the stars tend to be in a thicker configuration, either because of dynamical heating \citep{quinn_1993, abadi_2003, elmegreen_2006}, their formation from dynamically hot gas \citep{Comeron2014}, or stellar migration \citep{Halle2018}.
This bursty phase is then followed by a second, more stable phase in which a prominent thin disk is established.
Even though the thin disk forms stars at all times, the continuous star formation until recent times is the main driver for its younger, more metal-rich and less [Mg/Fe]-enhanced stellar populations.

\subsubsection{Numerical simulations}
\label{subsec:numerical_sim}

Numerical simulations have provided additional support to the scenarios discussed here concerning thin and thick disk formation. 
In most cases, they support a two-phase scenario in which an early fast phase would lead to the formation of a thick disk, followed by a slow phase assembling the thin disk at a low mass-assembly rate 
\citep{Dominguez-Tenreiro2017}. 
The first fast phase would result in a high-$\alpha$ component, and the slower second phase in a low-$\alpha$ disk \citep{Grand2018}. 
Since our results are characterized by relatively young thick disks with extended star formation histories, the thick/thin disk dichotomy is less clear, which can be explained by an early phase that is extended instead of fast and short \citep{Grand2018}. 
Additionally, cosmological simulations have shown that thick disks are mainly formed in-situ, but they host a fraction of accreted stars that can be significant \citep{Park2021, pinna_2024}. 
Merger histories (but also other galaxy properties such as galaxy mass) have a strong impact on the formation and growth of thick disks, explaining the observed variety of thick disk properties \citep{pinna_2024, Yi2024, Pillepich2024}.
\\
In particular, \citet{pinna_2024} produced edge-on mock stellar-population maps of 24 simulated Milky Way-mass galaxies from the AURIGA project \citep{Grand2017}, which can be qualitatively compared to integral-field spectroscopy observations of edge-on galaxies. 
Within the diversity shown in the stellar-population properties, two main families of thick disks were identified at redshift $z$~=~0: the old, metal-poor and $\alpha$-enhanced thick disks, and the younger thick disks with no sharp chemical distinction with the corresponding thin disks (see, e.g., Fig.~3 in \citealt{pinna_2024}). 
Analyzing different snapshots from high to low redshift (Fig.~10 and 11 in \citealt{pinna_2024}) shows that these thick disks had very different evolutionary paths, especially at the early stages. 
For similar masses at $z$~=~0, galaxies with a slow early evolution only reach low masses at high redshift (e.g., having about 10\% of the final mass at $z\sim 2$), which makes them more sensitive to later contributions, resulting in relatively young thick disks. 
These later contributions usually consist of the direct accretion of stars through mergers and the late star formation fueled by gas accretion. 
This type of evolutionary path tends as well to lead to a later formation of a dynamically cold thin disk. 
In contrast, galaxies with a rapid evolution at early times form their thick disks quickly at high redshift, and subsequently start to form a small primordial thinner and more metal-rich disk. 
The latter will keep developing slowly inside-out, while the thick disk will remain largely unchanged, preserving its old ages, low metallicities, and [Mg/Fe] enhancement from its formation.

\subsubsection{Metal-poor regions}
The metal-poor regions of the two galaxies ESO\,157-49 and IC\,1553 (Sect.~\ref{subsec:SFH}) show very low metallicities (lower than $-1$~dex) and the one of IC\,1553 is enhanced in [Mg/Fe] abundance (around 0.2~dex higher) compared to other regions in the thin disks. 
Furthermore, these metal-poor regions match spatially with regions of very high H$\alpha$ emission in \citet{rautio_2022} and \citet{gonzalez_2024}, and could indicate intense ongoing star formation from metal-poor gas.
However, we do not find stars that are significantly younger in this region than in the surroundings.
We tested if the age-metallicity degeneracy could be responsible for \textsc{ppxf} interpreting very young stars as being extremely metal-poor (Appendix~\ref{appendix:templates}). 
This was done by plotting SSP templates of different ages and metallicities (Fig.~\ref{fig:templates}) and comparing the strength of different age- and metallicity-sensitive absorption lines.
From this, we see that younger ages and lower metallicities have a similar impact on the H$\beta$ line (making the absorption more intense) and the magnesium and iron lines (making them less intense).
However, differences between the observed spectra of the metal-poor region and the spectra of a region with higher metallicity suggest that the stars have different properties in these regions.
We also see in Fig.~\ref{fig:spectra} that the magnesium and iron lines of the observed spectra in the metal-poor region are less intense than for observed spectra in a comparable region with higher metallicity, while the age-sensitive H$\beta$ absorption line has a similar intensity for both regions.
This suggests that those stars are, in fact, more metal-poor, and \textsc{ppxf} is interpreting the difference between spectra correctly as a difference in metallicity. 
Nevertheless, the intense H$\alpha$ emission in these regions suggests that the stars in the metal-poor regions are also very young. We cannot recover these small age differences with our method since these metal-poor regions are surrounded by stars that are already young, and age differences might be smaller or of the order of age uncertainties.

\subsubsection{ESO\,544-27}
\citet{rautio_2022} proposed that ESO\,544-27 has relatively older stellar populations than the other four galaxies with well-defined thin and thick disks, as this galaxy falls into the green valley and shows lower H$\alpha$ intensities. 
However, these small differences at very young ages (below a hundred Myr) are within our uncertainties and are not clear in our maps. 
\citet{Somawanshi2024} extracted mass-weighted stellar populations of ESO\,544-27 with a similar method to what is presented here.
Nevertheless, they used a different setup for \textsc{ppxf}, a different SSP library, and a different approach for the extinction correction, leading to large differences in this correction. 
Moreover, while we fit the emission lines of ESO\,544-27, these got masked in their study.
Moreover, with a S/N-threshold of 3, they cover mostly the morphological thin disk, since the transition to the thick disk happens at about 4~arcsec from the midplane according to the definition we used here (from \citealt{comeron_2018}, see Sect.~\ref{sub:SB_profiles}).
\citet{Somawanshi2024} also used a different definition of thin and thick disks, based solely on their own [$\alpha$/Fe] map. 

They found remarkably old mean ages (older than 9.8~Gyr) for a galaxy with current star formation. 
These old ages are uniform in the galaxy, with no clear differences at various distances from the midplane.
In fact, stars on the midplane look even slightly older ($\geq$~10~Gyr) than off-plane ($\sim$~9.95~Gyr) in their age map. 
Even if there is a bias of \textsc{ppxf} toward very old ages, we do not find in our work Voronoi bins with mean stellar ages around 10~Gyr in this galaxy. 
As mentioned above, all galaxies in our sample show mean ages below 8~Gyr for the majority of Voronoi bins. 
While \citet{Somawanshi2024} did not find a clear vertical metallicity gradient, in a region that corresponds mostly to our thin disk, their values in that same region they analyzed are similar to what we obtained. 
Contrary to us, they found a clear [$\alpha$/Fe] dichotomy, with $\alpha$ enhancement at larger distances from the midplane, and low [$\alpha$/Fe] in the midplane region. 
This was used to define a low-[$\alpha$/Fe] thin disk and high-[$\alpha$/Fe] outer thick disk. 
The above-mentioned differences in the method could be responsible for the discrepancies between their results and our results (e.g., the age maps). 
Regarding the star formation histories, their results are compatible with ours. They detected a young (< 2~Gyr) metal-rich stellar population mostly present in the outer thick disk (Fig.~9 in \citealt{Somawanshi2024}), supporting an in-situ formation for both thick and thin disks.
The star formation was nearly quenched between 2 to 8~Gyr and reignited around 1~Gyr ago in the thick disk,  and very recently ($\sim$~600~Myr ago) in the thin disk, probably by a wet merger event as proposed by \citet{Somawanshi2024}.

\subsection{Thin and thick disk masses}
\label{subsec:discuss_mass}
All galaxies with distinct thin and thick disks show a larger mass fraction in the thin disk dominated region (around 61 to 77~\%) than in the thick disk dominated region (around 22 to 39~\%), as shown in Tab.~\ref{tab:masses}. 
\citet{comeron_2018} calculated thin and thick disk masses 
(respectively, $M_\text{t}$ and $M_\text{T}$ in Tab.~\ref{tab:prop}).
$M_\text{t}$ are around 59 to 68~\% and $M_\text{T}$ are 32 to 41~\% of the total galaxy stellar mass. 
This discrepancy in the mass fractions comes from the different methods and the different wavelength ranges that were used. 
While we used the optical range, where a larger amount of light is obscured by the dust, \citet{comeron_2018} used images in the infrared.
This wavelength range is less affected by the dust than the MUSE wavelength range, and this affects mostly the mass of the thin disk, where the dust is located. 
More importantly, we calculated our mass fractions from restricted regions covered by the Voronoi bins since the MUSE pointings did not cover the full galaxies. 
The Spitzer S$^4$G \citep{sheth_2010} images used in \citet{comeron_2018}, on the other hand, cover the entire galaxies. 
In particular, we possibly cover a smaller fraction of the faint thick disk than S$^4$G data, resulting in a smaller thick disk mass fraction and a larger thin disk fraction. 
Further, \citet{comeron_2018} assumed a fixed mass-to-light ratio for the thin and the thick disk, while our mass-to-light ratios vary for different Voronoi bins (with different combinations of SSP models). 
Even if this is a second-order effect, the choice of a mass-to-light ratio has an important impact on the mass estimates.
Another reason why the predictions are different is that all the light above a certain height is assigned to the thick disk, and all the light below it is assigned to the thin disk. 
But at all heights, there is a contribution from both disks. 
So even though the thin disk dominates the surface brightness close to the midplane, a significant fraction of the thick disk total mass is hiding there. 
While we calculate thick and thin disk masses only in the region where they dominate, \citet{comeron_2018} used mass profiles that extend, for each disk, to the full vertical coverage of the data. 
Therefore, the thick disk mass fractions provided in our work are underestimated.
Despite this discrepancy in mass to \citet{comeron_2018}, the thin disks in the sample clearly host a much higher mass than the thick disk (higher by almost a factor of two, according to our results and \citealt{comeron_2018}).
As most of the galaxies have circular velocities $\geq$~90~km~s$^{-1}$, this distribution of thin and thick disk masses corresponds well with the findings of \citet{yoachim_2006} mentioned in Sect.~\ref{sec:introduction}, where they predicted that thick disks are usually less massive than thin disks.

\subsection{Galaxies with one dominating disk}
The galaxies with only one dominating disk component (especially ESO\,443-21 and PGC\,30591) may show slightly different star formation histories from galaxies with two distinct components.
The intermediate aged stellar population component (see Sect.~\ref{subsec:discuss_formation}) might not be as pronounced, and they rather show a more continuous decrease along those ages (around 10 to 6~Gyr).
If we take the uncertainty bands into account, these galaxies may also show an increase in star formation around intermediate ages.
Moreover, these three galaxies show a recent star-forming burst at very young ages (younger than 4~Gyr), similar to thin disks in the sample. 
These results are in agreement with \citeauthor{zaragoza-cardiel_2020} (\citeyear{zaragoza-cardiel_2020}, see also Sect.~\ref{sec:sample}), who detected star formation in the last Gyr for PGC\,28308, ESO\,443-21 and PGC\,30591.
This suggests a similar recent formation history for these gas-rich and young galaxies, where gas accretion is probably playing a role.
In fact, previous studies showed signs of interactions and gas accretion in some galaxies.
PGC\,28308 has disturbed gaseous outskirts and forms a pair with MCG-02-25-019 (see Sect.~\ref{sec:sample}, \citealt{Tully_2009, Meyer_2004}).
So, dynamical interactions with the paired galaxy might be responsible for the disturbances and the recent increase in star formation.
These disturbances might also have brought existing stars into dynamically hotter orbits, contributing to the mass assembly of the thick disk component.
Also, ESO\,443-21 was classified by \citet{schwarzkopf_2000} as an interacting/merging candidate.
Together with a negative radial metallicity gradient (Fig.~\ref{fig:lw_maps}), we suggest that this galaxy is growing a thin disk structure from the inside-out that might become more pronounced in the future.
This is similar to NGC\,3501, where the presence of a metal-rich inner thin disk in the midplane suggested an ongoing inside-out formation of a future prominent thin disk \citep{sattler_2023}.
However, we can only speculate about the growth of a prominent future thin disk, and the development of distinct thin and thick disks during the evolution might not hold for all galaxies.
On the other hand, these single-disk galaxies may already have two distinct disk components in reality, which are just not detected because of a slightly lower inclination angle (not perfectly edge-on) compared to the other galaxies.
Especially in low-mass galaxies, the absence of well-defined and thin dust lanes makes it difficult to ensure an edge-on orientation of the galaxy.
So some of these single-disk galaxies might not show well-defined thick and thin disks because they are not perfectly edge-on (see Fig.~9 in \citealt{comeron_2011}). 
This would explain why differences in the stellar populations are displayed between the midplane region and regions at larger heights.

\section{Summary and conclusion}
\label{sec:conclusion}

We used the MUSE data of eight late-type galaxies from \citet{comeron_2019} to derive their stellar populations (age, metallicity, and [Mg/Fe] abundance) and star formation histories. 
This is the first systematic stellar population mapping of a sample of edge-on very late-type galaxies that are gas-rich and have extraplanar ionized gas. 
In this type of galaxy, we find younger ages and lower metallicities in the thick and thin disks than were previously found in earlier-type disk galaxies.
While thick disks do not show the typical $\alpha$-enhancement that was formerly seen in other edge-on galaxies, they are systematically younger than was found previously, with ages that were traditionally associated with thin disks in earlier-type galaxies \citep[e.g.,][]{pinna_2019b, martig_2021}. 
The differences between thick and thin disks in our sample are rather small and suggest a slower and more continuous upside-down formation than in previously studied earlier-type galaxies.
\\
Our star formation histories show that a phase of intense star formation (potentially with minor mergers and gas accretion) that lasted until about 7 to 8~Gyr might support the formation of a thick disk.
We detected a variety of intermediate and older stars in the star formation histories of the thick disks, which might suggest that this intense phase heated stars from the midplane and left a variety of ages in thicker orbits.
This was observed in the stellar population maps, where the thick disks tend to be more metal-poor and more enhanced in the [Mg/Fe] abundance than the thin disks. 
For the thin disks in this sample, the star formation histories showed an increase in the mass fraction of the youngest stars compared to intermediate ages. This does not appear in the thick disks.
We therefore suggest that after an intense and likely turbulent star formation phase, a steadier phase begins in which the thin disk is established by continuous and efficient star formation that is probably fueled by gas accretion.
\\
In galaxies in which only one disk component dominates, we did not see a pronounced intermediate-aged stellar population, but rather a more continuous decrease at about 10 to 6~Gyr.
Similar to the galaxies with well-defined thin and thick disks, the single-disk galaxies also had star formation bursts at very young ages.
We proposed that ongoing star formation, together with dynamical interactions in these galaxies, might form more enhanced thin disk components in the future.
We also discussed, however, that a far from edge-on inclination could hide an already established two-disk structure in these galaxies.
It does not have to be universally true either that all galaxies eventually develop distinct thin and thick disks during their evolution.
\\\\
To conclude, this work is the first attempt to unveil the formation and evolution of thin and thick disks in very young dusty and gaseous edge-on spiral galaxies based on a spatially resolved stellar population analysis.
However, more work needs to be done in this field to gain a better understanding of the thick  and thin disk formation in spiral galaxies.
In addition, while testing other SSP models and fitting codes helps us to better understand the current systematics, improvements such as a higher S/N and a wider wavelength coverage would further enhance the reliability of the results.
In the future, larger surveys of edge-on galaxies such as GECKOS\footnote{\url{https://geckos-survey.org/index.html}} \citep{vandeSande2023} will help us to determine the leading formation mechanisms of thick and thin disks.

\begin{acknowledgements}
      NS acknowledges support from the Deutsche Forschungsgemeinschaft (DFG, German Research Foundation) -- Project-ID 138713538 -- SFB 881 (``The Milky Way System'', subproject B08) and from the Deutsche Forschungsgemeinschaft (DFG, German Research Foundation) in the form of an Emmy Noether Research Group (grant number KR4598/2-1, PI Kreckel) and the European Research Council’s starting grant ERC StG-101077573 (“ISM-METALS").
      FP acknowledges support from the Agencia Estatal de Investigación del Ministerio de Ciencia e Innovación (MCIN/AEI/ 10.13039/501100011033) under grant (PID2021-128131NB-I00) and the European Regional Development Fund (ERDF) "A way of making Europe”. FP acknowledges support also from the Horizon Europe research and innovation programme under the Marie Skłodowska-Curie grant “TraNSLate” No 101108180. 
      MM acknowledges support from the UK Science and Technology Facilities Council through grant ST/Y002490/1.
      IMN and JFB acknowledge support from the PID2022-140869NB-I00 grant from the Spanish Ministry of Science and Innovation.
      
\end{acknowledgements}

   \bibliographystyle{aa} 
   \bibliography{8gal.bib}

\begin{appendix}
\appendix
\counterwithin{figure}{section}

\section{Age-metallicity degeneracy}
\label{appendix:templates}

\begin{figure*}[b]
\centering
    \includegraphics[width=0.93\textwidth]{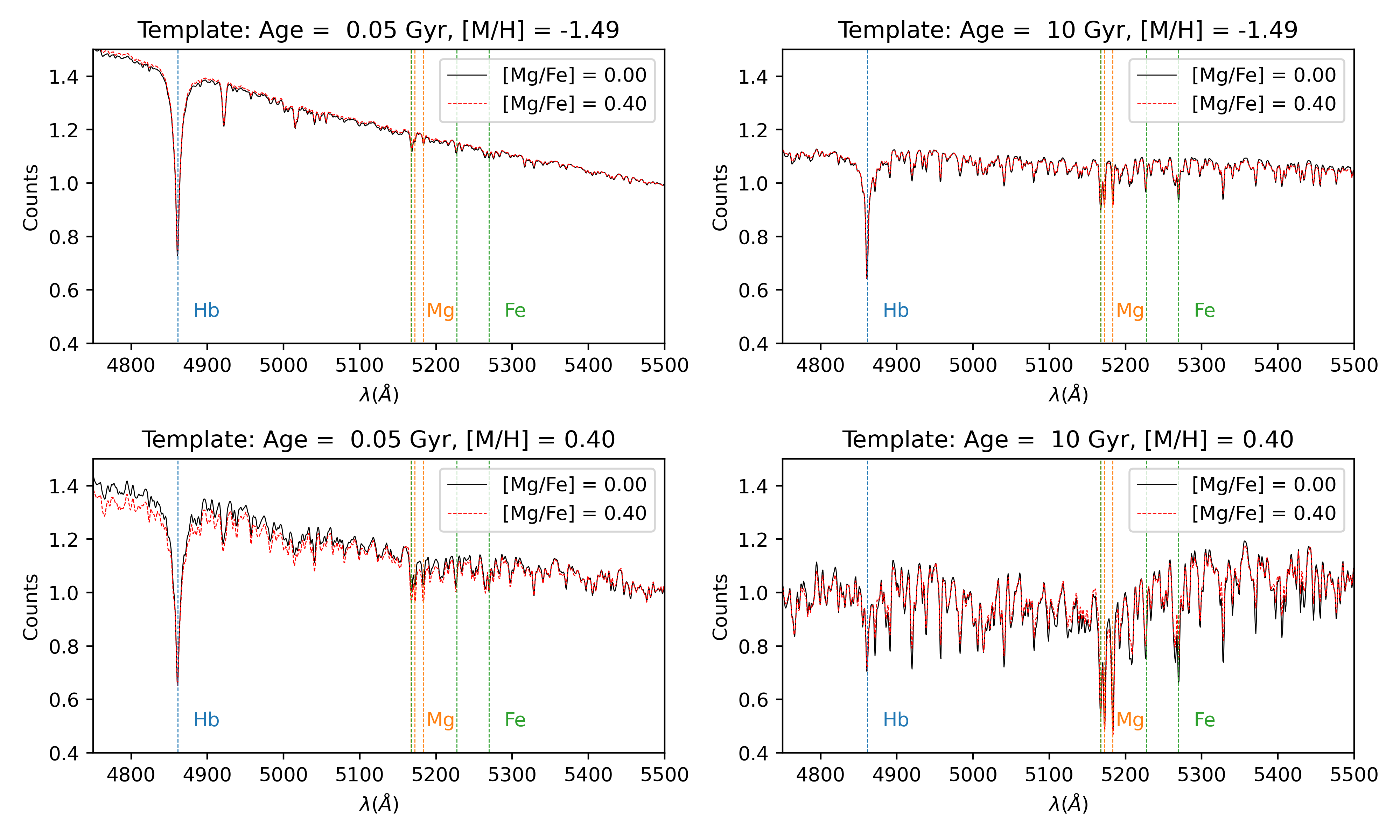}
    \caption{Comparison of MILES SSP templates with different ages (0.05 and 10~Gyr, from left to right) and metallicities (-1.49 and 0.40~dex, from top to bottom). The templates are shown for both values of [Mg/Fe] (black: 0.00~dex, red: 0.40~dex) and important absorption features are indicated by colored dashed lines (blue: H$\beta$ at 4861.33~$\si{\angstrom}$, orange: Mg at 5183, 5172 and 5167~$\si{\angstrom}$, green: Fe at 5167.49, 5227.15 and 5269.54~$\si{\angstrom}$).}
    \label{fig:templates}
\end{figure*}

In this Appendix, we show the tests on the age-metallicity degeneracy of the stellar templates discussed in Sect.~\ref{sec:discussion}.
Fig.~\ref{fig:templates} shows a comparison of MILES SSP models having different ages, metallicities, and [Mg/Fe] abundances.

\begin{figure*}
\centering
    \includegraphics[width=0.93\textwidth]{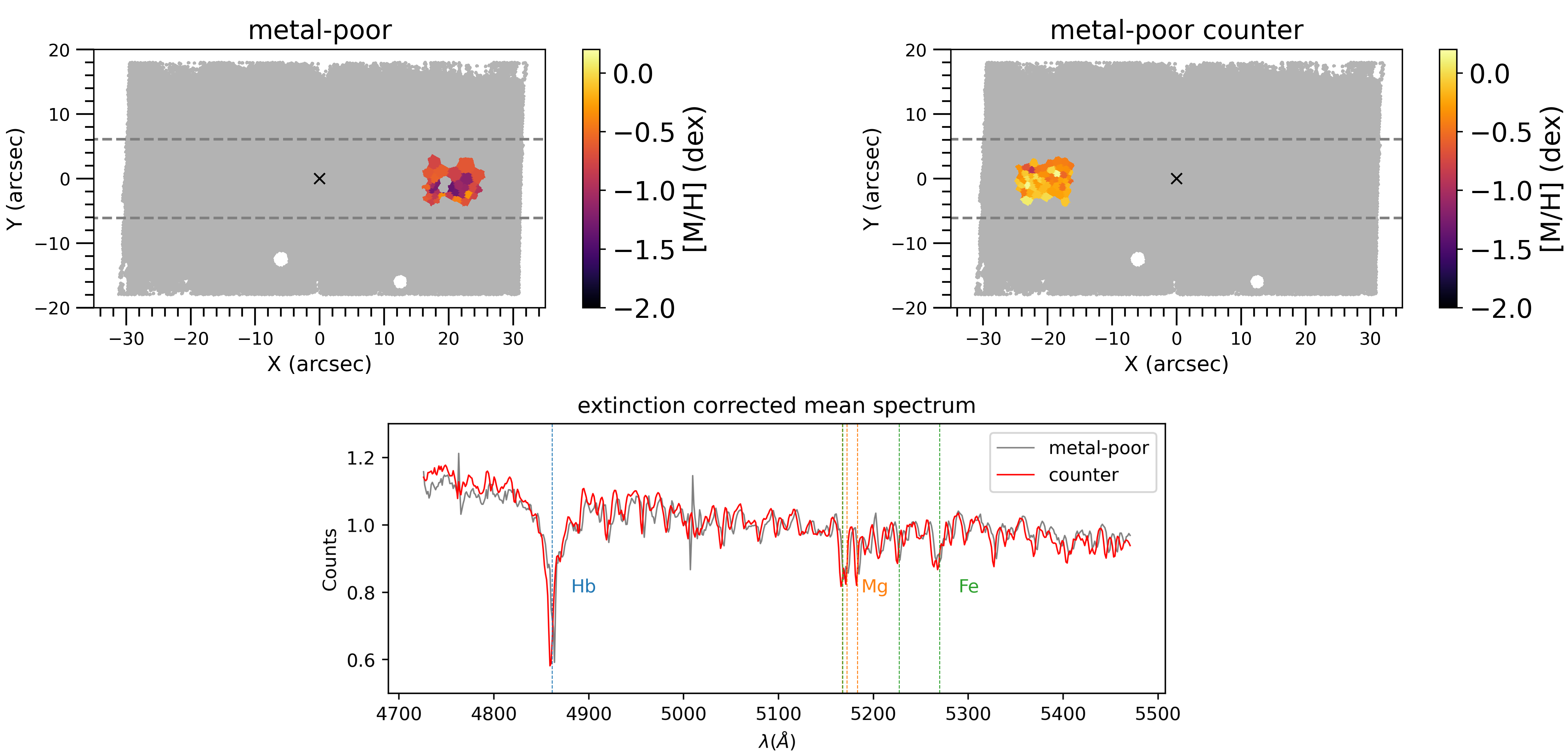}
    \caption{Comparison of the extinction corrected mean spectra of the metal-poor region (gray) and a counterpart region (red) on the other side of the center for ESO\,157-49. }
    \label{fig:spectra}
\end{figure*}

The two panels at the top show templates with different ages (0.05~Gyr and 10~Gyr) and the same sub-solar [M/H] (-1.49~dex), whereas the bottom panels show ages (0.05~Gyr and 10~Gyr) and super-solar [M/H] (0.40~dex).
Templates having the same ages (0.05~Gyr) and different [M/H] (-1.49~dex and 0.40~dex) can be compared using the two left panels, whereas the same comparison can be done on older ages (10~Gyr) using the two right panels.
Additionally, all panels show the two corresponding templates with [Mg/Fe]~=~0.00 and 0.40 as black solid and red dashed curves, respectively. 
Comparing all the different SSP models, one can see that younger ages and lower metallicities have a similar impact on the H$\beta$ line (making the absorption more intense) and the Mg and Fe lines (making them less intense).
In Fig.~\ref{fig:spectra}, we further show the differences in the spectra of the metal-poor region compared to a region, at a similar radius but on the opposite side of the disk, with higher metallicity.\\
It can be observed that the mean spectrum of the metal-poor region has less intense magnesium and iron lines than the counterpart mean spectrum, while the H$\beta$ line seems to be quite similar.\\
These differences in the spectra suggest that \textsc{ppxf} is interpreting the line intensities correctly as a difference in metallicity.
From intense H$\alpha$ emission \citep{rautio_2022} in these regions, we can further conclude that the stars in the metal-poor regions are also very young.
However, recovering the age differences between the metal-poor region and its surroundings is not possible with our method, as all ages in the thin disk are very young, and the age differences between these regions might be smaller than the age resolution or the uncertainties.
Nevertheless, when computing stellar properties, the age-metallicity degeneracy should always be taken into account, and results need to be handled with caution.

\newpage

\section{Tests for stellar population fitting}
\label{appendix:pop}
\begin{figure*}
	\centering
		\includegraphics[width=1\textwidth]{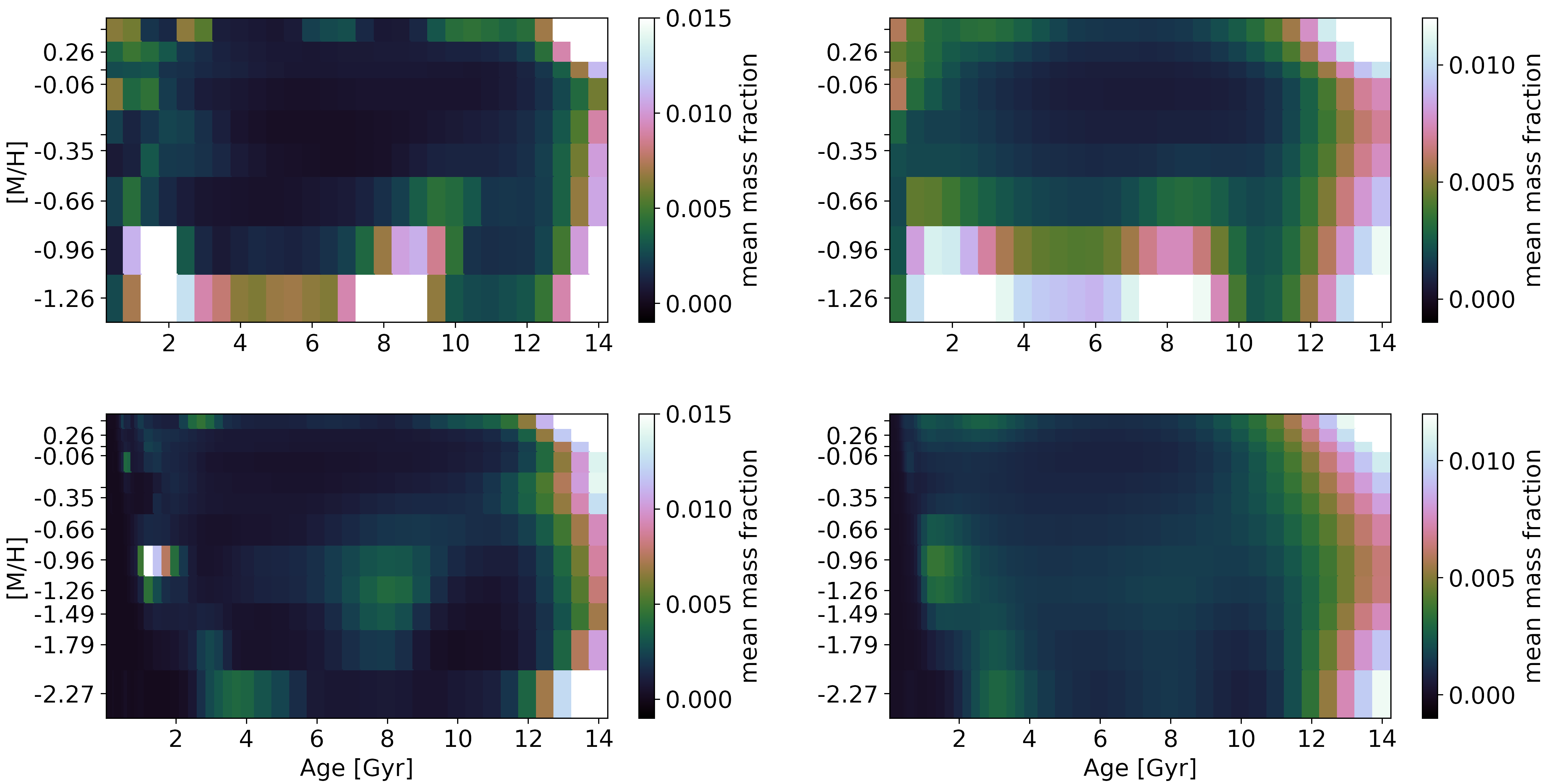}
        \caption{Star formation history tests for ESO\,443-21. In a metallicity-age grid, we show for this Voronoi bin the color-coded mass fraction assigned to different ages and metallicities summed up for both values of [Mg/Fe] (=~0.0 and 0.40~dex)
        The top figures show the results using limited models and a regularization~=~1~(left)~/~10~(right).
        The bottom figures show the results for the full range of models and a regularization~=~1~(left)~/~10~(right).}
	\label{fig:SFH_tests}
\end{figure*}

For the fitting of the stellar populations, we tested two different MILES SSP model selections:
\begin{itemize}
    \item Limited set of models (as in \citealt{sattler_2023}) that are in the safe range\footnote{\url{http://research.iac.es/proyecto/miles/pages/ssp-models/safe-ranges.php}} and provide the same resolution in the full age range:\\ $-1.26$~dex $\leq$ metallicity $\leq$ $+0.40$~dex, 0.5~Gyr $\leq$ age $\leq$ 14~Gyr evenly spaced every 0.5~Gyr, [Mg/Fe] = 0.0 and 0.4~dex
    \item Full range of models, with varying age resolution depending on the age range:\\ $-2.27$~dex $\leq$ metallicity $\leq$ $+0.40$~dex, 0.03~Gyr $\leq$ age $\leq$ 14~Gyr, [Mg/Fe] = 0.0 and 0.4~dex
\end{itemize}
together with the two different values of regularization~=~1 and 10. 
In Fig.~\ref{fig:SFH_tests}, examples of the star formation history grids resulting from these tests are shown for ESO\,443-21.

One can see that the limited models and regularization~=~1 fit show peaks in star formation at 14~Gyr (bimodal in [M/H]), 8.5~Gyr ([M/H]~$\approx$~-1.26~dex) and 2~Gyr ([M/H]~$\approx$~-0.96~dex). 
These peaks are smoother for the limited set of models and regularization~=~10 fit, but remain around the same ages and metallicity.
Using the full range of models, SSPs of lower metallicities are selected for the best fit by \textsc{ppxf}. 
This suggests that, when using the restricted set of models (with [M/H]~$\geq$~-1.26~dex) \textsc{ppxf} is hitting the boundaries of low metallicities.
This explains why when the [M/H]~$\leq$~1.49~dex SSP models are added, they are used by \textsc{ppxf} for the fitting.
Therefore, we need these lower metallicity models to properly fit the metal-poor peaks. 
Also, the star formation history is overall smoother in the bottom panels (full set of models) than the top panels (limited model grid) for the same regularization, and the 8.5~Gyr peak is less pronounced but still noticeable.
For the full range of models and regularization~=~10, the star formation peaks are even smoother, and the 8.5~Gyr peak is very hard to detect. 
So the same regularization has a stronger effect when we use this set of models. 
Finally, we decided to use the full range of MILES SSP models and regularization~=~1 for the full sample.

\newpage

\section{Tests for star formation histories}
\subsection{Recovery of age features}
\label{appendix:SFHmock}

\begin{figure*}[b!]
    \centering
    \includegraphics[width=\textwidth]{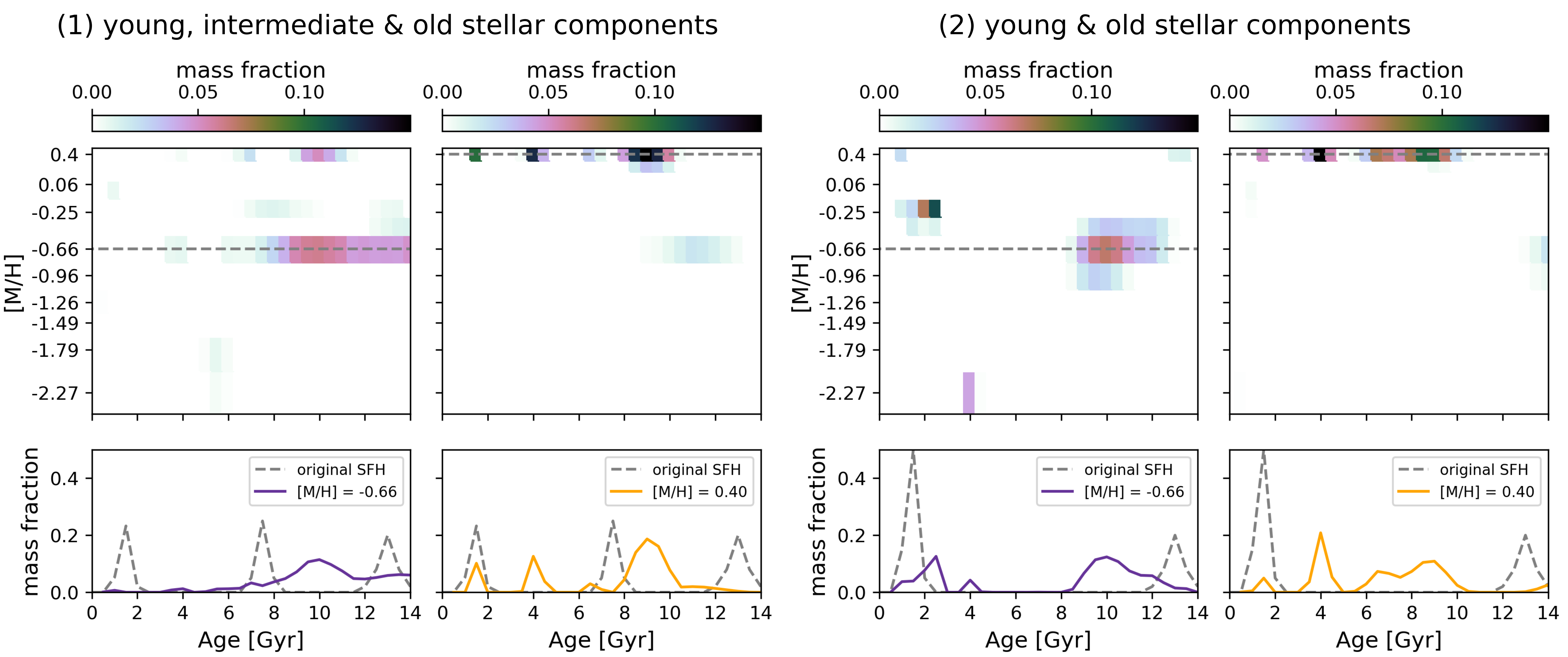}
    \caption{Star formation histories for two example mock distributions from the tests described in Sect.~\ref{appendix:SFHmock}.
    On the left, we show the results of mock spectra obtained by combining SSP models of young, intermediate, and old ages, while the right panels show results of mock spectra obtained by combining only young and old ages.
    For each of the two initial age distributions, a two-dimensional grid of metallicity and age is shown in the top panels for the tested metallicities [M/H] = -0.66 and 0.40~dex from left to right, which is indicated with a dashed gray line.
    In these grid plots, the mass fraction that \textsc{ppxf} gives to each SSP template is color-coded.
    The bottom panels show the star formation history distribution that is recovered from these top grids (solid lines) in terms of mass fraction per age, together with the initial star formation history (SFH) age distribution from which the mock spectra were created (dashed gray line).}
    \label{fig:SFH_tests_mock}
\end{figure*}

To test the limitations of our fitting method, we tested the \textsc{ppxf} fitting on mock spectra, following \citet{pinna_2019b}. We produced mock spectra with 11 different initial star formation histories to check if their features were recovered throughout the fitting process.
These initial star formation histories have peaks at young, intermediate, and old ages, with varying mass fractions for each of them.
For each of the initial star formation histories, mock spectra were created separately for three different values of metallicity: [M/H]~=~-0.66, 0.06, and 0.40~dex (low, solar, and high metallicities). 
We used ''base'' MILES SSP models \citep{vazdekis_2015}, which assume a solar [Mg/Fe] abundance at high metallicities, and a [Mg/Fe] enhancement at low metallicities.
Artificial noise was added to each of the mock spectra to reach a S/N of 60, as this is used as target-S/N in most of our sample galaxies (see Sect.~\ref{subsec:binning}).
Then, all noisy spectra were fit with \textsc{ppxf} using the same setup used for our galaxies (see Sect.~\ref{sec:methods}), including a regularization parameter of 1. 
The resulting weight distributions as well as the recovered star formation histories for some example cases can be seen in Fig.~\ref{fig:SFH_tests_mock}. 
In the left panels, we show a more complex star formation history with a young, intermediate, and old population, and in the right panels, a combination of young and old populations. 
As we are also interested in the recovery of metallicity and in the age-metallicity degeneracy, the weight distributions and star formation histories are presented for metal-poor and metal-rich stellar populations. 
To get a better and more robust understanding of the fitting consistency, we also performed Monte Carlo simulations, similar to the ones described in Sect.~\ref{subsec:MC_simulations}, with 100 realizations and regularization of 0 for each mock spectrum. 
In general, the average results of the 100 Monte Carlo realizations with regularization~=~0, show a similar behavior as the single fits with regularization~=~1.
\\
These tests on the mock spectra show that the initial star formation history peaks are mostly recovered, but often smoothed and at different age positions. 
Errors in the age positions are often large, of up to 4~Gyr, especially for intermediate and old ages, as it was suggested by our estimates of uncertainties from Monte Carlo simulations (Sect.~\ref{subsec:res_SFH}). 
Initial peaks of young ages are recovered quite well, especially for mock spectra with solar and high metallicity.
However, for the more complex star formation history and the lowest metallicity, the mass fraction assigned by \textsc{ppxf} to the young component is very small (left panel in Fig.~\ref{fig:SFH_tests_mock}). 
For peaks at intermediate ages, \textsc{ppxf} fits them with lower-aged SSP models (around 3~Gyr younger than the initial peak) while also assigning different metallicities than the original one, showing that age-metallicity degeneracy (App.~\ref{appendix:templates}) definitely plays a role here.
Furthermore, some mass fraction is assigned to very old stars for low metallicities (to a smaller extent for solar metallicities) even when this component is not present in the input star formation history (not shown). 
However, surprisingly, star formation history peaks occurring at old ages are often preferably fit by \textsc{ppxf} with SSP models that are around 4~Gyr younger than the original age, similar to the intermediate-aged peaks. 
Also, the large error bars in age for the old initial star formation history peaks lead to the appearance of an intermediate aged component, when it was absent in the original slope, especially for the metal-rich mock spectra (see right panels in Fig.~\ref{fig:SFH_tests_mock}). 
However, these mock star formation histories are taken as simple example cases, but they are unrealistic. 
Real galaxies would very unlikely form most of their mass fraction in only one, two, or three short bursts. 
While these tests are very useful to better understand the limitations of our method, they must be considered as unrealistically worst cases showing upper limits in our uncertainties. 
Further testing of realistic star formation histories is beyond the scope of this paper. 
Moreover, as the galaxies in our sample were, in general, metal-poor, and the mentioned effects become more important for higher metallicities, we consider our conclusions to still hold within the estimated uncertainties.
\\
These tests suggest that the reader rely more on a qualitative interpretation of star formation histories derived from \textsc{ppxf}, referring to ages qualitatively as young, intermediate, or old, rather than relying on absolute age values. 
On the other hand, metallicity is in general well recovered, especially for the spectra with [M/H]~=~$-0.66$ and 0.40~dex.
For the nearly solar metallicity of 0.06~dex, SSP models with higher and lower metallicity are combined in the fitting process to obtain an average best-fit solar metallicity. 
\\
Compared to the tests conducted by \citet{pinna_2019b}, we obtained larger errors in the position of peaks in the star formation histories. 
This can be explained by the increased complexity of our tests. 
The mock star formation histories analyzed here are more complex and include multiple stellar population components spanning a range of ages, rather than being dominated solely by old populations as in \citet{pinna_2019b}. 
In particular, young and intermediate-aged populations are included, whose SSP spectra tend to be less accurate. 
Additionally, we did not use the same set of models to generate the mock spectra and to fit them.
While the mock spectra were created using base MILES SSP models, the fits were performed using models with varying [Mg/Fe] enhancements. 
In contrast, \citet{pinna_2019b} used the same set of MILES SSP models, with two different [Mg/Fe] enhancements, for both generating and fitting the mock spectra with \textsc{ppxf}.
To conclude, the recovery of the star formation history remains challenging due to various factors, including systematics in the models, the limitations of spectral coverage, and the intrinsic assumptions in the fitting process. 
In the context of these tests, the input star formation histories are significantly more extreme than the peaks retrieved from the actual data. 
Since \textsc{ppxf} is not designed to work with sparse solutions, the imposed regularization naturally prevents such extreme variations and features.
As a result, while the star formation histories in Fig.~\ref{fig:SFH_tests_mock} do not fully reflect reality, this does not imply that the results obtained in this study are similarly inaccurate.
Additionally, the S/N and the lack of bluer optical coverage limit the precision of age determinations.
Despite these challenges, our analysis provides valuable insights, and we anticipate future methodological improvements that will enhance the reliability of star formation history recoveries. 
These will consist of improvements in the SSP models and the complementary use of instruments covering additional age-sensitive features in the bluer wavelength range of spectra.

\newpage

\subsection{Varying thin disk scale heights}
\label{appendix:SFHthinheight}

\begin{figure*}[b!]
    \centering
    \includegraphics[width=0.8\textwidth]{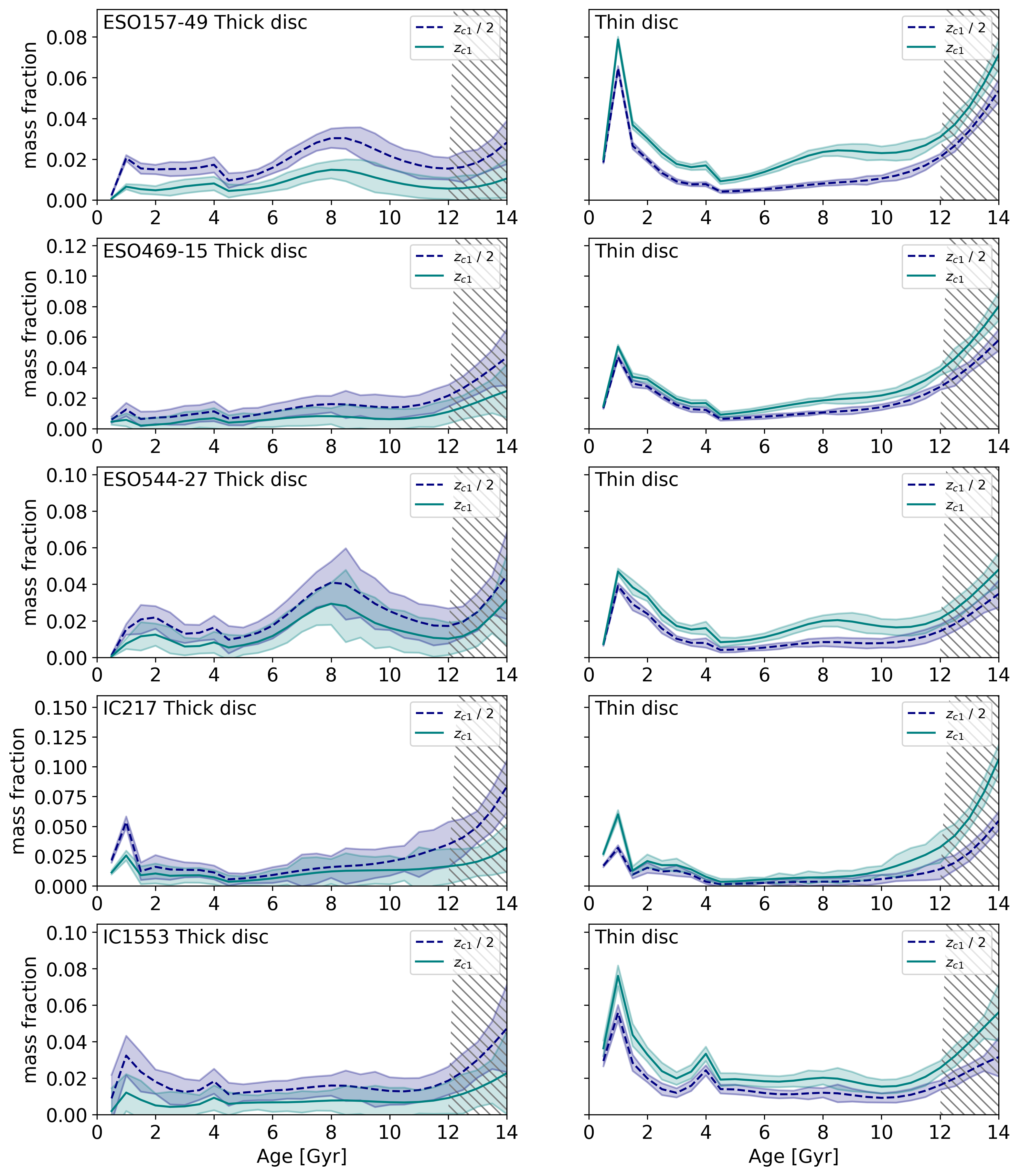}
    \caption{Component decomposition tests of the star formation histories in terms of mass fraction per age with evenly spaced age bins every 0.5~Gyr. 
    The left and right columns show the star formation histories for thick and thin disks, respectively, for the five out of the eight galaxies containing two well-defined disk components.
    In each panel, the light blue solid line represents the component decomposition as in Fig.~\ref{fig:SFH_maps} with a adapted thin disk scale height of $z_{c1}$, whereas the dark blue dashed line represents a adapted thin disk scale height of $z_{c1}$/2.
    Uncertainties from Monte Carlo simulations are shown by the colored shaded areas.
    Large mass fractions at old ages should be handled with caution, as there might be a bias of \textsc{ppxf} fitting toward very old ages (see Sect.~\ref{subsec:res_SFH}) and are shaded with a gray area.}
    \label{fig:SFH_maps_test1}
\end{figure*}
To test the robustness of the star formation histories under changes in the thick and thin disk identification, we performed the same analysis as described in Sect.~\ref{subsec:SFH} but with a narrower thin disk scale height of $z_{c1}$/2 (see Tab.~\ref{tab:prop} for values of $z_{c1}$).
We chose this value of the photometric scale height as a clean determination of the height scales using stellar population parameters might be arbitrary, and is not possible here.
However, we only want to test here the qualitative changes of varying the thin disk height on the star formation histories.
Also, with a thin disk scale height of $z_{c1}$/2 we are only including the youngest and most metal-rich Voronoi bins in the thin disk (except for IC\,217 as its thin disk is metal-poorer as the thick disk, see Sect.~\ref{subsec:res_pop}).
\\
The results of this star formation history test can be seen in Fig.~\ref{fig:SFH_maps_test1}, where mass fractions for each age are presented for the thin and thick disks of individual galaxies, similar to Fig.~\ref{fig:SFH_maps}.
This test shows that all thick disks host, in general, more mass at all ages.
Especially, ESO\,157-49 shows higher mass fractions around the peak at intermediate and the youngest ages, while IC\,217 hosts more mass at the oldest and youngest ages.
The thin disks, however, host less mass in general, while ESO\,157-49 and ESO\,544-27 in particular show a more continuous decrease of star formation along intermediate ages.
The high populations of very young and old stars are still present along all thin disks, even though mass fractions are a bit lower compared to the higher thin disk scale height of $z_{c1}$.
\\
These shifts of more mass into the thick disks, less mass in the thin disks as well as lower mass fractions of intermediate aged stars in the thin disks were expected since a lower thin disk scale height of $z_{c1}$/2 excludes more Voronoi bins with certain mass and young stars from the thin disk component.
Also, the stellar populations along intermediate ages seem to vanish from the thin disk star formation histories using narrower component definitions, as these are more distributed over the thick disks as well as the transition region between the thin and thick disks.
Moreover, a higher fraction of young stars in these more extended thick disks can be explained by a larger contamination of thin disk stars.
\\
Putting everything together, this test proves that the results of the star formation histories presented in Sect.~\ref{subsec:res_SFH} and Fig.~\ref{fig:SFH_maps} are robust against minor variations of a thin disk scale height.

\section{Mass-weighted stellar populations}
\label{appendix:mw}

\begin{figure*}
    \centering
    \includegraphics[width=0.8\textwidth]{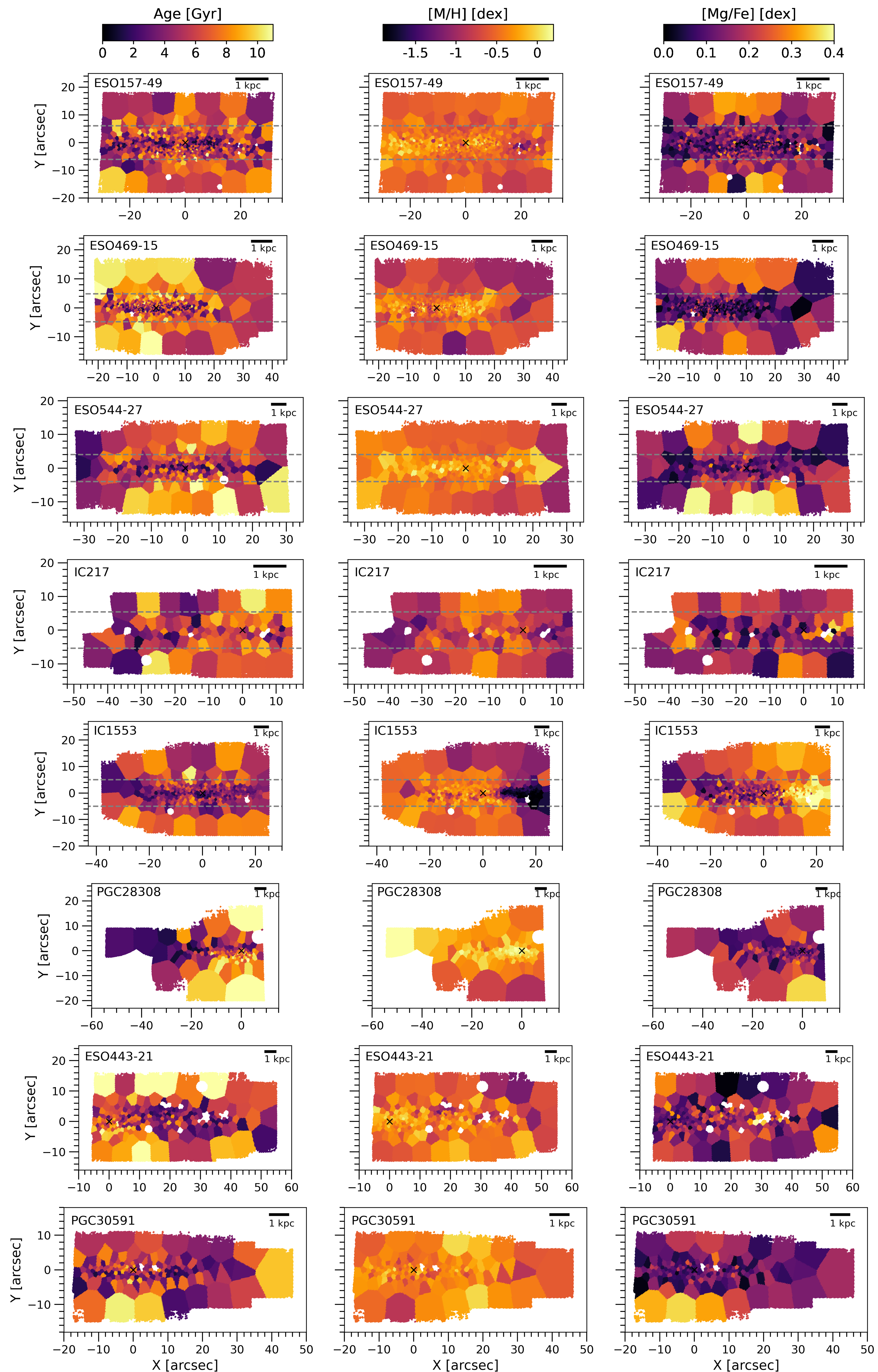}
    \caption{Mass-weighted age (left), metallicity (middle), and [Mg/Fe] abundance (right) maps for the full sample. The dashed gray lines mark the regions above and below which the thick disk dominates the vertical surface brightness profiles.}
    \label{fig:mw_maps}
\end{figure*}

Here we present the mass-weighted stellar population maps in Fig.~\ref{fig:mw_maps}.

Comparing the light- and mass-weighted stellar population maps, the mass-weighted ones show, on average, older ages, higher metallicities, and lower [Mg/Fe] abundances (also seen in \citealt{sanchez-blazquez_2014, sattler_2023}).
This difference in age is because younger stars contribute more to the light-weighted results due to their higher luminosity compared to older stars. 
In contrast, older stars contribute to a higher mass, leading to more prominence in mass-weighted results. 
As a consequence, the light-weighted map tends to exhibit younger average ages compared to the mass-weighted map.
Further, as indicated in Tab.~\ref{tab:uncertainties_mw}, the mass-weighted stellar population properties show larger uncertainties on average compared to the light-weighted ones.

\begin{table}[h!]
    \centering
    \caption{Average uncertainties for the mass-weighted stellar population maps calculated as the mean uncertainty of all bins belonging to the corresponding component.}
    \begin{tabular}{p{1.8cm} p{1.6cm} p{0.8cm} p{1cm} p{1.3cm}}
        \hline
		\\
		\hfill{} Galaxy & Component & $\Delta$Age & $\Delta$[M/H] & $\Delta$[Mg/Fe]\\
        \hfill{} \ & \ & [Gyr] & [dex] & [dex]\\
        \hline
        \\
		\hfill{} ESO\,157-49 & thin disk \newline thick disk & 0.95 \newline 1.99 & 0.09 \newline 0.12 & 0.13 \newline 0.15 \\
		\hfill{} ESO\,469-15  & thin disk \newline thick disk & 1.06 \newline 2.63 & 0.09 \newline 0.28 & 0.13 \newline 0.24 \\
        \hfill{} ESO\,544-27  & thin disk \newline thick disk & 1.03 \newline 1.87 & 0.09 \newline 0.15 & 0.10 \newline 0.15 \\
        \hfill{} IC\,217  & thin disk \newline thick disk & 1.05 \newline 2.03 & 0.14 \newline 0.14 & 0.14 \newline 0.17 \\
        \hfill{} IC\,1553  & thin disk \newline thick disk & 3.24 \newline 5.04 & 0.40 \newline 0.57 & 0.37 \newline 0.26 \\
        \hfill{} PGC\,28308  & full galaxy & 0.78 & 0.06 & 0.18 \\
        \hfill{} ESO\,443-21 & full galaxy & 1.40 & 0.15 & 0.22 \\
        \hfill{} PGC\,30591 & full galaxy & 1.00 & 0.10 & 0.13 \\
        \hline
		\\
    \end{tabular}
    \label{tab:uncertainties_mw}
\end{table}

\newpage
\ 
\newpage

\section{Uncertainties for light-weighted stellar populations}
\label{appendix:uncertainties}

\begin{figure*}
    \centering
    \includegraphics[width=0.8\textwidth]{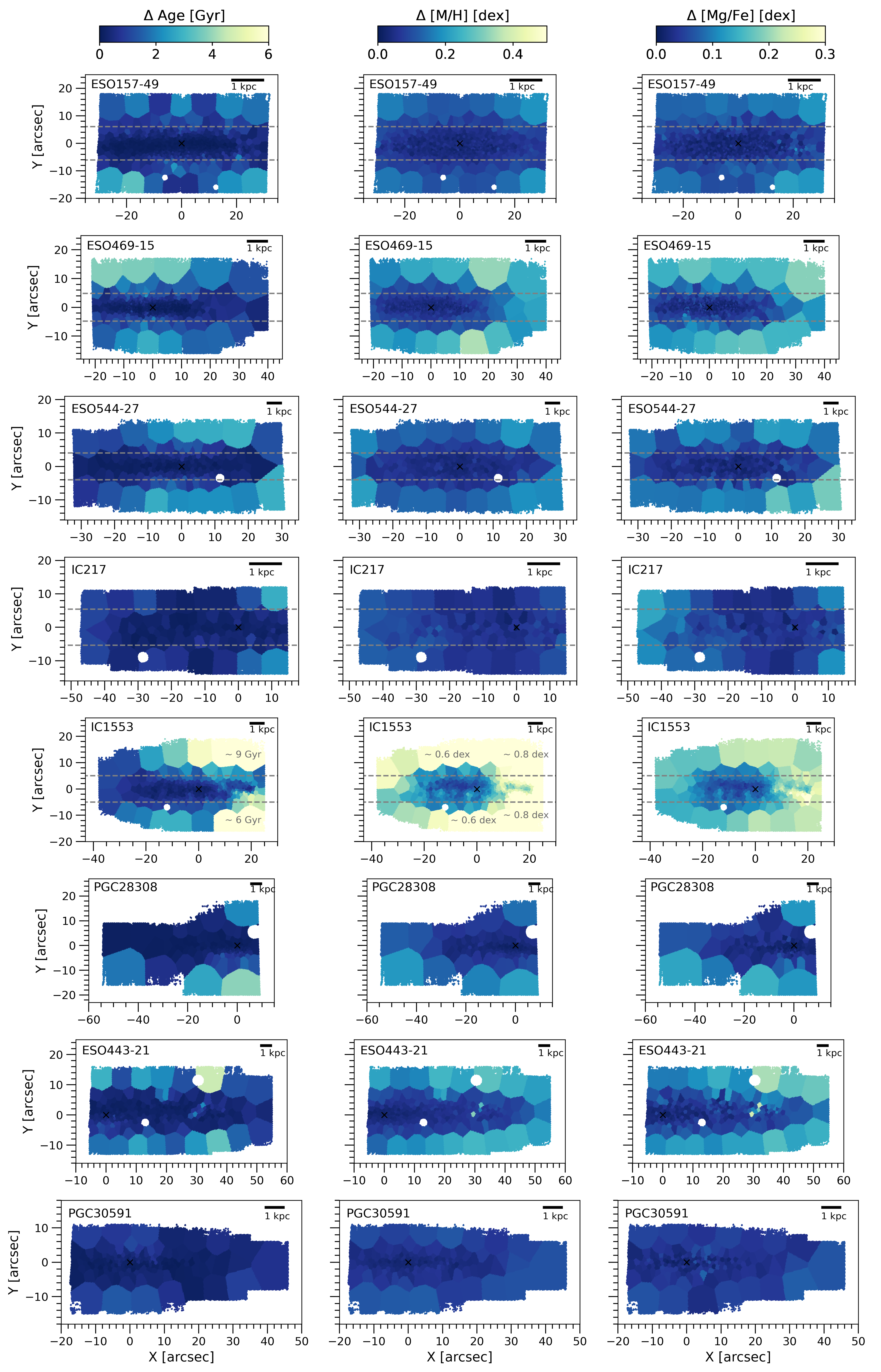}
    \caption{Uncertainties of the light-weighted age (left), metallicity (middle), and [Mg/Fe] abundance (right) maps for the full sample. 
    For IC\,1553, when the color scale saturates, we indicated the approximate uncertainty in the Voronoi bins by numbers.
    The dashed gray lines mark the regions above and below which the thick disk dominates the vertical surface brightness profiles.}
    \label{fig:MC_lw_maps}
\end{figure*}

In Fig.~\ref{fig:MC_lw_maps}~ we show the uncertainties of the light-weighted stellar populations as derived by Monte Carlo simulations described in Sect.~\ref{subsec:MC_simulations}.
In general, the uncertainties for all galaxies are smaller in the midplane region and get larger for regions farther away, which is explained by the lower S/N per Voronoi bin and also per pixel before even Voronoi binning these regions.
Tab.~\ref{tab:uncertainties_lw} contains a summary of the average uncertainties for each galaxy in light-weighted age, metallicity, and [Mg/Fe] abundances.

\end{appendix}
\end{document}